\documentclass{aa}  

\usepackage{subfig}
\usepackage{graphicx}
\usepackage{lscape}
\usepackage{hyperref}

\usepackage{txfonts}
\newcommand{\ms}{M$_{\odot}$}
\newcommand{\usfr}{M$_{\odot}$ yr$^{-1}$}

\newcommand{\kms}{km s$^{-1}$}
\newcommand{\Kkms}{K km s$^{-1}$}
\newcommand{\jykms}{Jy km s$^{-1}$}

\newcommand{\lco}{$L^{\prime}_{\rm CO(1-0)}$}
\newcommand{\aco}{$\alpha_{\rm CO}$}

\newcommand{\uaco}{$M_{\odot}$ pc$^{-2}$ (K km s$^{-1}$)$^{-1}$}
\newcommand{\uLco}{K km s$^{-1}$ pc$^2$}
\newcommand{\as}{$^{\prime\prime}$}
\newcommand{\sirs}{\emph{Spitzer}/IRS}
\newcommand{\tdep}{$t_{depl}$}
\newcommand{\mstar}{$M_{\star}$}
\newcommand{\mdust}{$M_{dust}$}
\newcommand{\mh}{$M_{H_2}$}
\newcommand{\lir}{$L_{IR}$}
\newcommand{\lpah}{$L_{PAH}$}

\begin{document} 

   \title{The molecular gas properties in local Seyfert 2 galaxies}
    \author{
    \href{https://orcid.org/0000-0003-4751-7421}{F. Salvestrini\inst{1}} 
    \and
    C. Gruppioni\inst{2,3}
    \and
    E. Hatziminaoglou\inst{4}
    \and
    F. Pozzi\inst{2,3}
    \and
    C. Vignali\inst{2,3}
    \and 
    V. Casasola\inst{5}
    \and 
    R. Paladino\inst{3,5}
    \and 
    S. Aalto\inst{7}
    \and
    P. Andreani\inst{4}
    \and 
    S. Marchesi\inst{2,6}
    \and 
    T. Stanke\inst{4}
    }

   \institute{
    INAF – Osservatorio Astrofisico di Arcetri, Largo E. Fermi 5, 50125, Firenze, Italy
   \and
   Dipartimento di Astronomia ``Augusto Righi'', Universit\`a degli Studi di Bologna, Viale Carlo Berti Pichat 6/2, 40127, Bologna, Italy
   \and
   INAF - Osservatorio di Astrofisica e Scienza dello Spazio di Bologna, Via Gobetti 93/3 - 40129 Bologna - Italy
   \and
   ESO, Karl-Schwarzschild-Str 2, D-85748 Garching bei M\"unchen, Germany
   \and
   INAF - Istituto di Radioastronomia, Via P. Gobetti 101, 40129 Bologna, Italy
    \and
   Department of Physics and Astronomy, Clemson University,  Kinard Lab of Physics, Clemson, SC 29634, USA
   \and
   Department of Space, Earth and Environment Chalmers University of Technology, SE-412 96 Gothenburg, Sweden
    \and
   Italian ALMA Regional Centre, Via P. Gobetti 101, 40129, Bologna, Italy
     }



\abstract {} 
{We present a multi-wavelength study of the molecular gas properties of a sample of local Seyfert 2 galaxies to assess if, and to what extent, the presence of an active galactic nucleus (AGN) can affect the Interstellar Medium (ISM) properties in a sample of 33 local Seyfert 2 galaxies.} 
{We compare the molecular gas content (M$_{H_2}$), derived from new and archival low-J CO line measurements of a sample of AGN and a control sample of star-forming galaxies (SFGs).
Both the AGN and the control sample are characterised in terms of host-galaxy properties (e.g., stellar and dust masses, \mstar\ and \mdust, respectively; and star formation rate, SFR).
We also investigate the effect of AGN activity onto the emission of polycyclic aromatic hydrocarbon (PAH) molecules in the mid-infrared (MIR), a waveband where the dust-reprocessed emission from the obscured AGN contributes the most.}
{The AGN hosted in less massive galaxies (i.e., $M_{\star}<10^{10.5}$ M$_{\odot}$; $M_{dust}<10^{7.5}$ M$_{\odot}$) show larger molecular gas content with respect to SFGs matched in stellar and dust mass.
When comparing their depletion time ($t_{dep}\propto M_{H_2}/SFR$), AGN show $t_{dep}\sim0.3-1.0$ Gyr, similar to those observed in the control sample of SFGs.\\
Seyfert 2 galaxies show fainter PAH luminosity at increasingly larger dominance of the nuclear activity in the MIR.
} 
{We find no clear evidence for a systematic reduction of the molecular gas reservoir at galactic scale in Seyfert galaxies with respect to SFGs.
This is in agreement with recent studies showing that molecular gas content only is reduced in regions of sub-kpc size, where the emission from the accreting supermassive black hole dominates.
Nonetheless, we show that the impact of AGN activity on the ISM is clearly visible as suppression of the PAH luminosity.}

   \maketitle
%
%
\section{Introduction}
The star formation (SF) activity in galaxies can be affected by the presence of an accreting supermassive black hole (SMBH).
Indeed, active galactic nuclei (AGN) have been accounted for both suppressing and enhancing the SF activity in their host galaxy (e.g., \citealt{Feruglio10}; \citealt{CanoDiaz12}; \citealt{Cicone14}; \citealt{Carniani15}; \citealt{Fiore17}; \citealt{CresciMaiolino18}), through negative and positive feedback, respectively.
For instance, by injecting a large amount of energy in the circumnuclear region, the AGN may prevent the gravitational collapse of molecular clouds, hence the formation of new stars, in processes referred to as ``negative'' feedback (e.g., \citealt{Ellison21}).\\
In recent years, a multitude of studies have been dedicated to the investigation of the physical properties of the molecular gas in galaxies, aimed at understanding the effect of nuclear activity on the gas kinematics (feeding and feedback; e.g., \citealt{GarciaBurillo03}; \citealt{Combes13}; \citealt{Fluetsch19}; \citealt{FernandezOntiveros20}) or the driving mechanism of the excitation of the molecular component (e.g., \citealt{Daddi15}; \citealt{Pozzi17}; \citealt{Mingozzi18}; \citealt{Leroy21a}; \citealt{Esposito22} and reference therein).
Interferometric facilities like ALMA and NOEMA provide a powerful tool for spatially resolved studies of the nuclear region, looking for AGN-driven outflows in active galaxies (e.g. \citealt{Combes13}; \citealt{Cicone14}; \citealt{GarciaBurillo14}; \citealt{Fiore17}; \citealt{Alonso-Herrero18}), while single-dish observations are needed to recover the large-scale emission of the host galaxy (e.g., \citealt{Leroy09}; \citealt{Saintonge11b}; \citealt{Saintonge17}; \citealt{JimenezDonaire19}; \citealt{Sorai19}).\\
When comparing the properties of active and inactive galaxies, a major issue is the uncertain determination of the level to which the presence of an AGN affects the star-formation rate (SFR) proxies in the far-infrared, FIR, as well in the ultraviolet, UV.
To address this issue, a multi-wavelength approach is crucial to disentangle the relative contribution from AGN and host-galaxy SF in the global output of the galaxy.\\
In recent years, our group developed a multi-wavelength strategy aimed at characterizing the emission from the AGN in a statistical sample of local Seyfert galaxies.
This method requires the collection of observations in different bands (from the X-rays to the mm band) and their coherent analysis to provide a complete picture of the interplay between the AGN and the host-galaxy SF activity. 
A previous work from our group \citep[][hereafter, G16]{gruppioni16} exploited the collection of photometric measurements -- from the UV to the FIR, including in particular \sirs\ mid-infrared (MIR) spectra -- to perform a detailed spectral energy distribution (SED) decomposition for a sample of 76 Seyfert galaxies.
The 76 objects are MIR-selected objects drawn from the active galaxies of the extended 12 micron galaxy sample (12MGS; \citealt{RMS93}) upon the availability of the \sirs\ spectra. 
G16 derived an almost complete characterization of the sources in terms of the relative contributions from stellar and nuclear activities to the global energy output of the galaxy.
Here, we extend the work by G16 by quantifying, when present, the impact of AGN activity on the cold molecular gas reservoir and MIR (3-25$\mu$m) emission in 33 Seyfert 2 galaxies out of the 76 objects in G16.
Our approach is twofold: $i$) we derive the molecular gas masses for the targets by using new observations with the Atacama Pathfinder EXperiment (APEX) antenna of the CO(2-1) emission line for 23 AGN, and by collecting low-J CO observations from the literature for the remaining 10 targets.
We then compared the molecular gas content and consumption time scales in this well-defined sample of obscured AGN with a control sample of star-forming galaxies (SFGs).
$ii$) We study the emission in the MIR to investigate the effects of the AGN radiation on ISM tracers in this band, as polycyclic aromatic hydrocarbon (PAH) features.\\
Several comparative studies of the molecular gas content and SF efficiency between local active galaxies and SFGs have been presented over the years (e.g., \citealt{Maiolino97}; \citealt{Bertram07}; \citealt{Rosario18}), suggesting that local AGN hosts do not differ from inactive sources, in contrast with the results observed in high-redshift quasars ($z\sim$1-3; e.g., \citealt{Brusa18}; \citealt{Talia18}).
While detailed studies are nowadays flourishing thanks to ALMA and NOEMA (e.g., \citealt{GarciaBurillo21}; \citealt{Leroy21a}), providing the insight on the physical conditions of the molecular component down to the size of the molecular clouds (few tens of pc), spatially-integrated studies remain powerful tools to investigate and compare the properties of active and inactive galaxies.
Galactic-scale studies are also important to compare the properties of local galaxies with the objects at higher redshift ($z>1$), where the high-spatially resolved studies on large samples are challenging.
In this context, we compare the molecular gas properties (namely, the molecular gas masses and depletion time) of a finely characterized sample of local Seyfert galaxies with a control sample of SFGs.
For the study of the MIR emission and among the wealth of diagnostics present in the MIR band (e.g., emission line ratios, both emission and absorption spectral features), we focused on the PAH features.
PAH features are usually associated with the presence of ongoing SF activity, and they have been proposed as an alternative to CO to trace the molecular gas content in SFGs \citep[][hereafter, Co19]{Cortzen19}.
However, it is widely debated how the presence of nuclear activity affects the PAH luminosity, since the high ionizing radiation field from the AGN is able to destroy PAH molecules.
Indeed, AGN have been observed to suppress the PAH emission \citep{Diamond-Stanic10}, especially the features at the shortest wavelength (i.e., at 6.2 and 7.7 $\mu$m).
On contrary, strong PAH emission was detected in local Seyfert galaxies (e.g., \citealt{Honig10}; \citealt{Alonso-Herrero14}), with recent works suggesting that AGN may enhance PAH features (e.g. \citealt{Jensen17}).
In this work, we investigate if the presence of the AGN affects the luminosity of PAH features with respect to a control sample of SFGs.
G16 provided the characterization of the AGN emission (namely, the bolometric luminosity and relative contribution to the global IR emission of the galaxy) in the sample of Seyfert 2 galaxies, which makes this sample a reference sample for this kind of study.\\
The paper is structured as follows: the sample is presented in Sect. \ref{sec:sample}; the CO spectra of the new APEX observations and archival CO spectroscopy, along which the multi-wavelength data included in the analysis, are presented in Sect. \ref{sec:data}.
We introduce the control sample of inactive galaxies used during the analysis in Sect. \ref{sec:control_sample}.
In Sect. \ref{sec:results} we analyzed the effect of AGN activity on the molecular gas masses and properties of the host galaxy, as well as the effect on the emission from MIR features.
The conclusion are drawn in Sect. \ref{sec:conclusion}.
In Appendix \ref{app:KS-test} we reported the results of the statistical test used to evaluate the difference between the properties of AGN and SFGs.\\
Throughout the paper, distance-dependent quantities are calculated for a standard flat $\Lambda CDM$ cosmology with the matter density parameter $\Omega_{M}=0.30$, the dark energy density parameter $\Omega_{\Lambda}=0.70,$ and the Hubble constant $H_0=70$ km s$^{-1}$ Mpc$^{-1}$ \citep{komatsu09}.
We adopt a solar oxygen abundance of $12+log(O/H)=8.69\pm0.05$ from \cite{Asplund09}. 
Errors are given at 68 per cent confidence level.

\begin{figure*}[hp]
	\includegraphics[width = \textwidth, keepaspectratio=True]{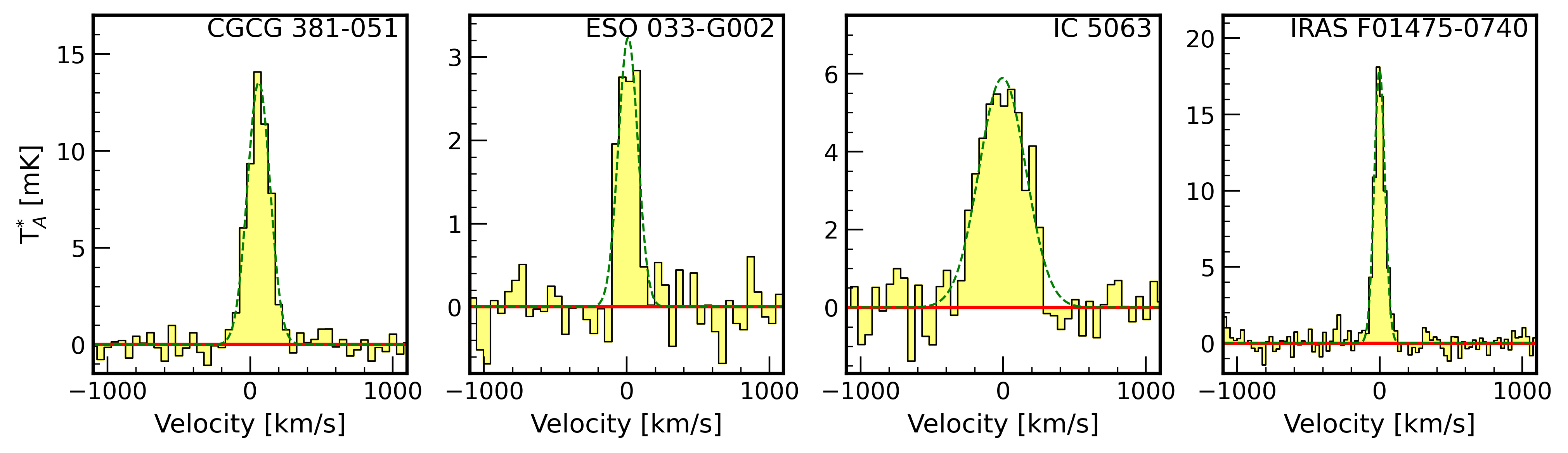}\\
	\includegraphics[width = \textwidth, keepaspectratio=True]{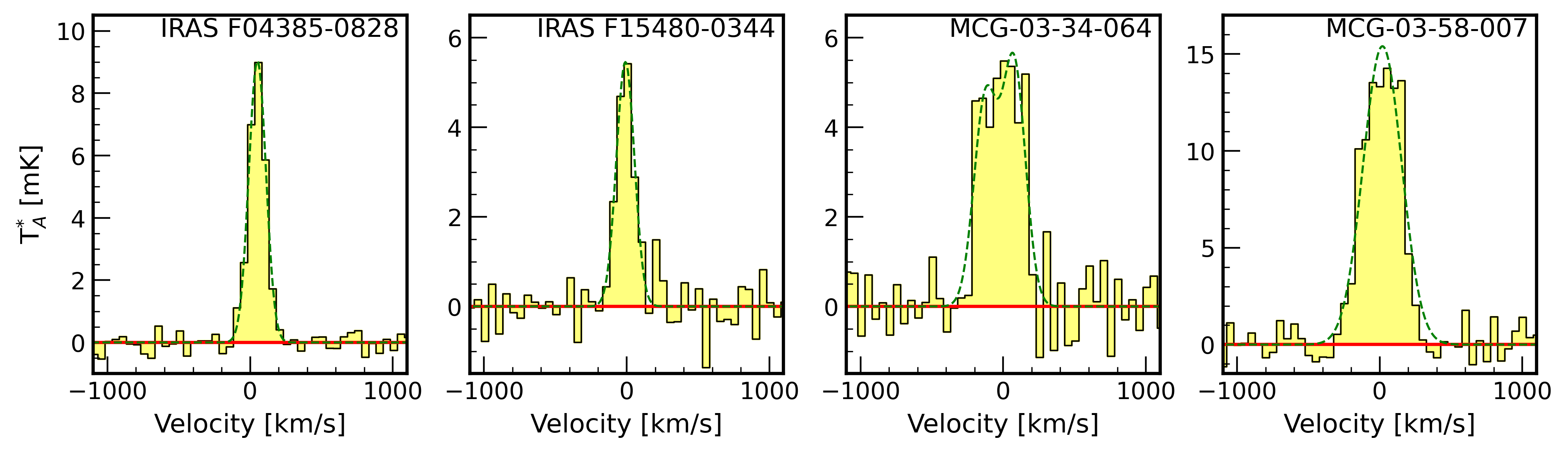}\\
	\includegraphics[width = \textwidth, keepaspectratio=True]{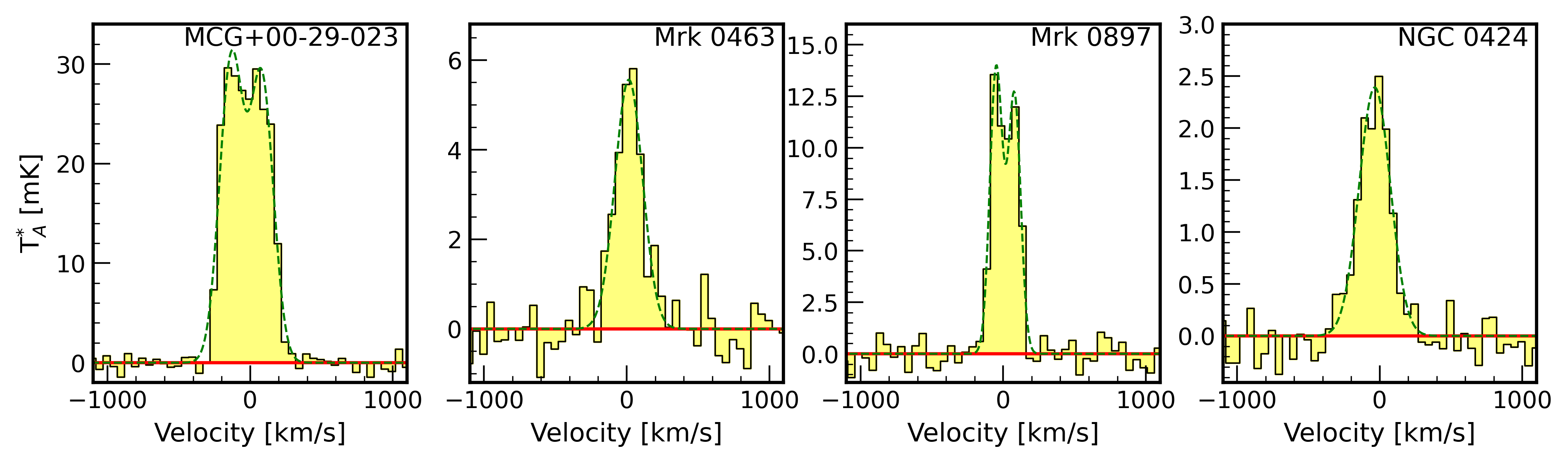}\\
	\includegraphics[width = \textwidth, keepaspectratio=True]{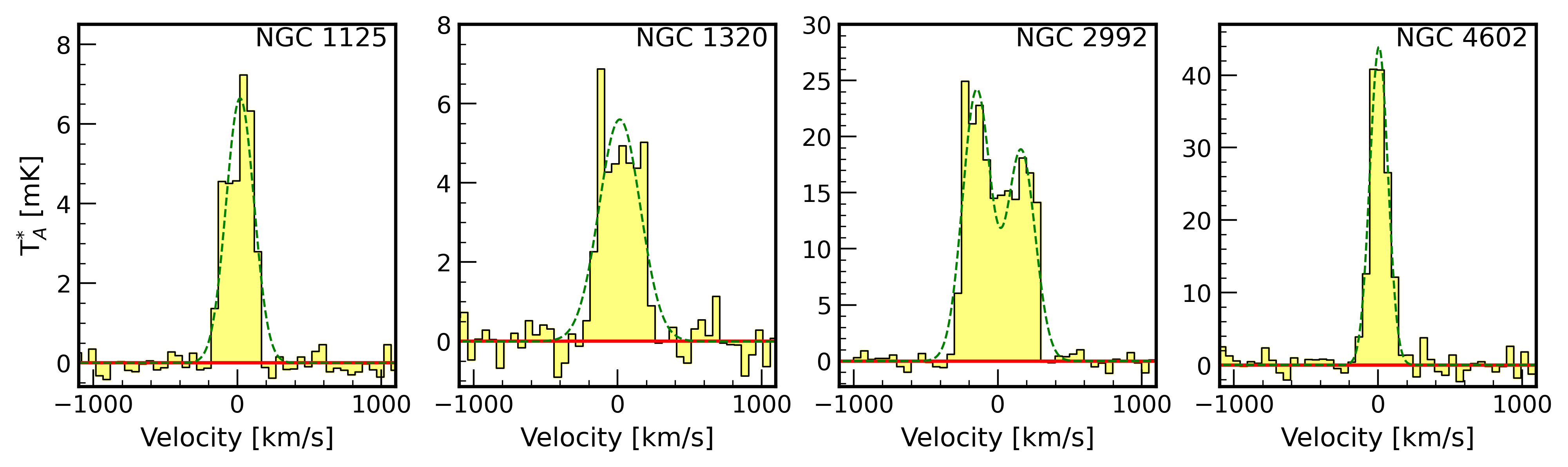}
	\caption{Continuum-subtracted CO(2-1) emission line profile for 23 Seyfert 2 galaxies observed by APEX. 
	Fluxes are expressed as antenna temperatures (T$_{A}^*$, in units of mK), while the spectral axis are in velocity units (\kms), calculated with respect to the CO(2-1) expected sky-frequency at the redshift of each source, assuming the radio conversion for the velocity.
	Each panel spans a fixed range of 2000 \kms of velocity around the systemic velocity of the galaxy to allow a simple visual comparison of the kinematics of the lines.
	The CO(2-1) emission is clearly detected in all observations, with at least S/N=4 in the channel corresponding to the peak of the line profile.
	The green-dashed line represents the best-fit function, that consists of one or two Gaussian functions, depending on the spectral line profile; the red solid line represents the zero K level.    
	}
	\label{fig:all_apex_spectra}
\end{figure*}
\begin{figure*}[ht] 
\ContinuedFloat
	\includegraphics[width = \textwidth, keepaspectratio=True]{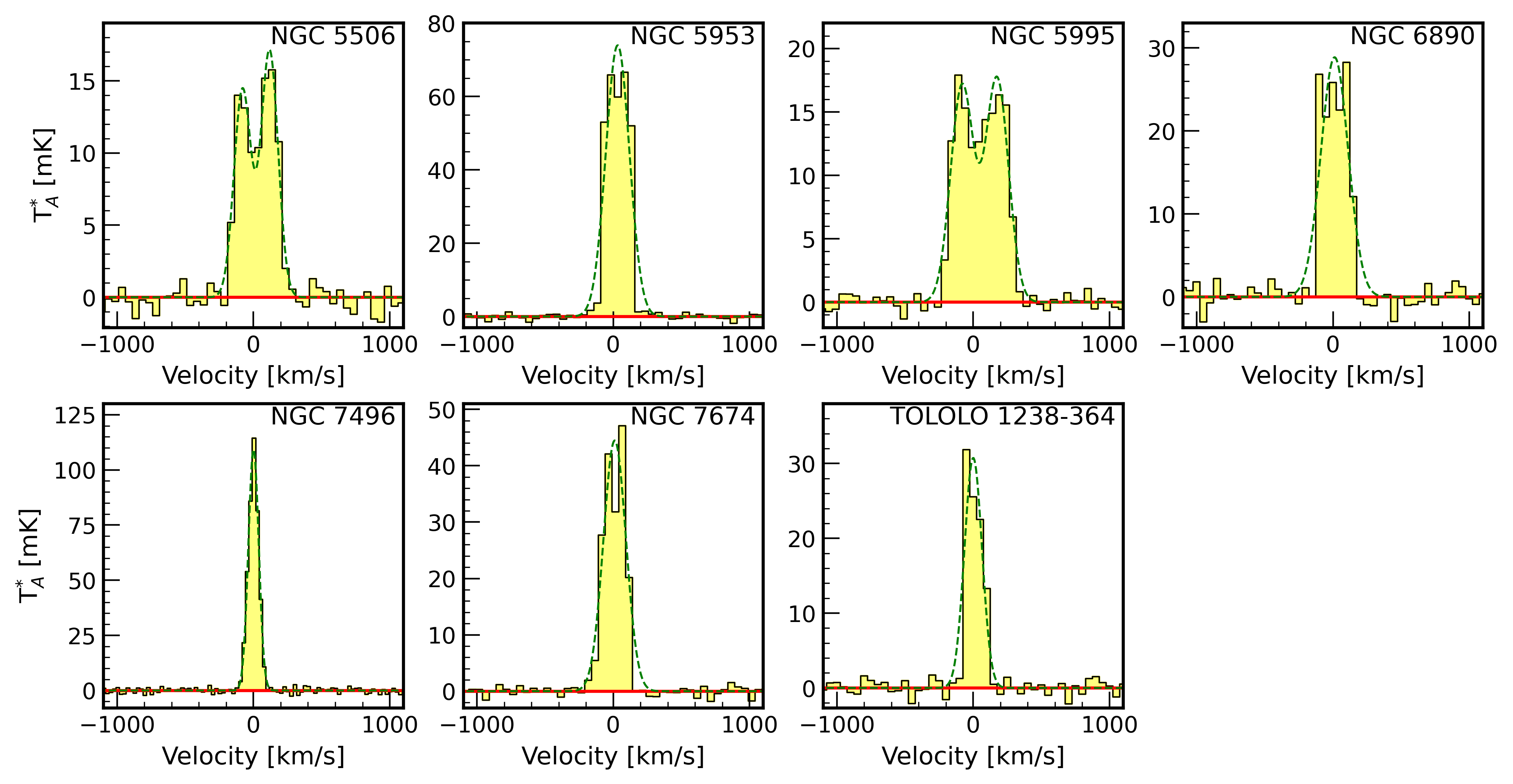}\\
	\caption{Continued.}
\end{figure*}

%
\section{The sample}
\label{sec:sample}
In the present work, we study the Seyfert 2 galaxies drawn from the sample of 76 Seyfert galaxies studied in G16.  
Among the 76 MIR-selected active galaxies presented in G16, we selected 33 optically-classified obscured sources with available CO spectroscopy from new and pre-existing observations (objects are listed in Table \ref{table:sample}).
The bulk of the sample consists of 23 objects for which we were granted 18h of observing time for CO(2-1) spectroscopy with the APEX telescope \citep{APEX}.
Ten more sources with low-J CO (J=1-0; 2-1) observations collected with several telescopes were added to the sample (references to the literature are reported in Table \ref{table:CO}).\\
The sources benefit from a detailed SED decomposition performed by G16, which provides the estimates of the SFR, stellar and dust mass of the targets, as well as the relative contribution from the AGN and host-galaxy to the total IR emission.
To compare the properties of the AGN with a control sample of SFGs, we need reliable characterization of the AGN host galaxies from the SED decomposition, including the emission of stars, dust heated by SF and AGN circumnuclear torus.
For this reason, we focused on Seyfert 2 galaxies, since the UV-optical is dominated by the stellar emission, making it easier to constrain the contribution from stars (hence, the stellar content, \mstar) with respect to type 1 AGN.
Indeed, when looking at Seyfert 2 galaxies, the AGN primary UV-optical emission, which arises from the accretion disk surrounding the SMBH, is blocked and absorbed by obscuring material, i.e. the circumnuclear dust, often assumed to have a toroidal shape, and is then re-emitted by the same dust in the MIR.\\

%
\section{Data}
\label{sec:data}
The multi-wavelength data used in this study consist of single-dish observations of CO emission lines to trace the molecular gas content, optical emission line intensities to determine the metal abundances, the results from the SED decomposition performed by G16, and the measurements of PAH features derived from the literature (\citealt{HernanCaballero11}; HC11, hereafter).
%
\subsection{Single-dish observations}
\subsubsection{APEX data reduction}
\label{sec:apex_reduction}
The observations of the CO(2-1) emission line (at 230.5 GHz rest-frame frequency) for 23 (out of 33) galaxies were carried out with the PI230 receiver (project 0103.F-9311, PI: F. Salvestrini), mounted on the APEX 12~m antenna.
The CO spectra were obtained with single-beam observations pointed at the optical positions of the targets, as provided by the NASA/IPAC Extragalactic Database (NED)\footnote{https://ned.ipac.caltech.edu/}. 
Since we were interested in the integrated line emission, we requested a spectral resolution of 50 \kms. 
The requested spectral resolution was sufficient to resolve the line profile with at least 6 channels assuming a Gaussian line profile with full width half maximum FWHM$\sim$300 \kms, as typically observed in the case of low-J transition in local active galaxies (e.g. \citealt{Papadopoulos12}).
Observations were designed to obtain a signal-to-noise ratio, S/N, of at least $\sim$6 at the peak of the line, corresponding to S/N$\sim$15 for the integrated line emission.
The resulting integration times on source ranged between few minutes up to a couple of hours, depending on the brightness of the source.
At the observing CO(2-1) frequency the average main beam size is $\theta_{ mb } = 27$\as, corresponding to physical scale of $\sim$10~kpc at median redshift ($z\sim$0.02) of the sample.\\
Data reduction was performed using the Continuum and Line Analysis Single-dish Software (CLASS), which is part of the GILDAS\footnote{\url{http://www.iram.fr/IRAMFR/GILDAS/}} software.
Calibrators were chosen according to the standard guideline for APEX observations\footnote{\url{http://www.apex-telescope.org/ns/apex-data/}}.
The CO(2-1) emission line profiles for the 23 sources are presented in Fig. \ref{fig:all_apex_spectra}.
We first fit the CO line emission with a single Gaussian profile, an approach that allows for a preliminary assessment of the central velocity ($\varv_{0}$), strength and width (meant as FWHM, $W_{CO}$) of the line.
Where the line profile show clear evidence of more than one peak, we repeated the fit with two Gaussian functions, as done for MCG-03-34-064, MCG+00-29-023, Mrk 0897, NGC 2992, NGC 5506, NGC 5995 (see Fig. \ref{fig:all_apex_spectra}).
In this case, we assumed as the central velocity and width of the CO line the mean of the central velocities and the squared sum of the FWHM of the two Gaussian functions, respectively.
However, neither the single and double Gaussian components allow us to properly model the CO emission, due to the complex profile and the low spectral resolution of the observations (i.e., $\delta \varv_C=50$ \kms).
This motivated our decision to estimate the total line fluxes by integrating the emission in a fixed velocity range [-1100, 1100] \kms centered on the systemic velocity of each source, once the baseline was subtracted.
This approach secured uniform and solid estimates of the line intensity, even in those cases where the observed line profile differs significantly from that of a single -- or double -- Gaussian function.
The error on the CO line fluxes were calculated as:
\begin{equation}
    \delta I_{CO} = \sigma_{RMS} (W_{CO} \delta \varv_C)^{1/2}, 
\end{equation}
where $\sigma_{rms}$ is the root-mean-square (RMS) noise in K (reported in Table \ref{table:CO}), $W_{CO}$ is the CO line width in \kms, and $\delta \varv_C$ is the spectral resolution ($\delta \varv_C=50$ \kms).
The RMS is calculated as the quadratic mean of the signal in the line-free channels, i.e. over the remaining side-bands (for a total of $\sim$ 2000 \kms) once the edge channels were flagged (3 channels per side).
Calibration uncertainties affect significantly the estimates of the CO line intensities, being larger than the spectral noise.
We conservatively assume them to be 10$\%$ of the intensity, as usually done for similar observations (e.g. \citealt{Csengeri16}; \citealt{Giannetti17}).
The uncertainties reported in Table \ref{table:CO} are the quadratic sum of the calibration uncertainties and the spectral noise integrated over the line profile.\\
To convert the line integrated intensities to fluxes in units of \jykms, we adopted a constant Jy/K conversion factor of 37$\pm$3, suitable for the PI230 receiver\footnote{http://www.apex-telescope.org/telescope/efficiency/}.
In Table \ref{table:CO}, we present the new CO fluxes, obtained with APEX.\\
%

\begin{sidewaystable*}

\begin{tabular}{ l l l l l l l l l l l}
\hline
Name & RA & DEC & D & D25 & $i$ & $\log(\frac{M_{\star}}{M_{\odot}})$ & $\log(\frac{M_{dust}}{M_{\odot}})$ & SFR & $12+\log(O/H)$ & f$_{AGN}$\\
(1) & (2) & (3) & (4) & (5) & (6) & (7) & (8) & (9) & (10)& (11)\\
\hline
CGCG 381-051       & 23h48m41.29s & +02d14m21.01s &135&1.848&34&10.43$\pm$0.08&8.11$\pm$0.09&13.26$\pm$0.07&8.92$\pm$0.14&0.01$\pm$0.42\\
ESO 033-G002       & 04h55m59.59s & -75d32m26.99s &79&1.971&30&10.60$\pm$0.08&8.36$\pm$0.08&2.86$\pm$0.02&9.19$\pm$0.21&0.61$\pm$0.02\\
IC 5063            & 20h52m1.99s & -57d4m9.01s &49&2.429&51&10.69$\pm$0.09&7.09$\pm$0.07&5.28$\pm$0.63&8.89$\pm$0.13&0.46$\pm$0.05\\
IRAS F01475-0740    & 01h50m2.69s & -07d25m48.0s &77&1.595&43&9.05$\pm$0.06&6.62$\pm$0.08&3.49$\pm$0.42&8.96$\pm$0.15&0.25$\pm$0.6\\
IRAS F04385-0828    & 04h40m54.91s & -08d22m22.01s &65&1.838&82&10.27$\pm$0.10&8.03$\pm$0.08&3.36$\pm$0.28&9.26$\pm$0.23&0.71$\pm$0.03\\
IRAS F15480-0344    & 15h50m41.50s & -03d53m17.99s &133&1.647&53&10.75$\pm$0.09&7.78$\pm$0.07&10.16$\pm$1.34&8.97$\pm$0.15&0.52$\pm$0.1\\
MCG-03-34-064       & 13h22m24.38s & -16d43m43.0s &72&2.129&57&10.65$\pm$0.09&7.25$\pm$0.08&5.65$\pm$0.56&9.60$\pm$0.30&0.72$\pm$0.07\\
MCG-03-58-007       & 22h49m36.91s & -19d16m23.99s &138&1.914&44&10.95$\pm$0.11&8.82$\pm$0.10&20.15$\pm$0.88&9.38$\pm$0.25&0.48$\pm$0.07\\
MCG+00-29-023       & 11h21m12.20s & -02d59m3.01s &109&1.92&42&10.95$\pm$0.11&8.82$\pm$0.09&18.71$\pm$0.55&9.05$\pm$0.17&0.34$\pm$0.13\\
Mrk 0273            & 13h44m42.11s & +55d53m12.65s &167&1.822&67&11.04$\pm$0.08&8.03$\pm$0.10&66.85$\pm$8.02&9.39$\pm$0.25&0.39$\pm$0.02\\
Mrk 0463            & 13h56m2.90s & +18d22m18.98s &224&2.035&59&11.22$\pm$0.07&7.45$\pm$0.11&15.4$\pm$2.03&8.81$\pm$0.10&0.87$\pm$0.02\\
Mrk 0897            & 21h7m45.80s & +03d52m40.01s &115&1.838&0&10.91$\pm$0.10&9.39$\pm$0.07&28.23$\pm$3.39&9.50$\pm$0.27&0.08$\pm$0.02\\
NGC 0034            & 00h11m06.55s & -12d06m26.33s &85&2.062&90&10.58$\pm$0.08&7.66$\pm$0.08&24.44$\pm$1.79&9.53$\pm$0.29&0.19$\pm$0.1\\
NGC 0424            & 01h11m27.49s & -38d5m1.0s &51&2.219&78&10.49$\pm$0.08&6.65$\pm$0.06&1.26$\pm$0.05&9.10$\pm$0.19&0.8$\pm$0.03\\
NGC 0513            & 01h24m26.85s & +33d47m58.01s &85&1.791&62&10.78$\pm$0.09&8.35$\pm$0.09&6.64$\pm$0.44&9.14$\pm$0.21&0.08$\pm$0.3\\
NGC 1125            & 02h51m40.39s & -16d39m1.98s &47&2.178&75&9.52$\pm$0.08&6.98$\pm$0.08&2.23$\pm$0.06&9.15$\pm$0.21&0.28$\pm$0.13\\
NGC 1320            & 03h24m48.71s & -03d2m33.0s &38&2.27&81&10.43$\pm$0.08&6.56$\pm$0.07&0.92$\pm$0.01&9.10$\pm$0.20&0.56$\pm$0.01\\
NGC 2992            & 09h45m42.01s & -14d19m35.0s &33&2.465&90&9.19$\pm$0.07&7.98$\pm$0.09&3.61$\pm$0.26&9.23$\pm$0.22&0.35$\pm$0.03\\
NGC 3079            & 10h01m57.80s & +55d40m47.24s &16&2.913&90&9.76$\pm$0.07&7.15$\pm$0.10&3.81$\pm$0.04&9.81$\pm$0.32&0.2$\pm$0.1\\
NGC 4388            & 12h25m46.75s & +12d39m43.51s &36&2.731&90&9.31$\pm$0.09&6.97$\pm$0.09&3.7$\pm$0.11&8.84$\pm$0.11&0.4$\pm$0.1\\
NGC 4602            & 12h40m36.52s & -5d7m54.98s &37&2.11&54&9.42$\pm$0.10&8.08$\pm$0.07&2.99$\pm$0.07&8.97$\pm$0.16&0.12$\pm$0.31\\
NGC 5135            & 13h25m44.06s & -29d50m01.2s &59&2.379&25&10.71$\pm$0.10&7.91$\pm$0.10&15.61$\pm$1.87&9.60$\pm$0.30&0.25$\pm$0.04\\
NGC 5256            & 13h38m17.50s & +48d16m37.0s &122&2.08$^{a}$&                 &10.42$\pm$0.08&8.26$\pm$0.07&31.72$\pm$1.39&9.16$\pm$0.2&0.14$\pm$0.17\\
NGC 5347            & 13h53m17.83s & +33d29m26.98s &34&2.21&45&10.11$\pm$0.11&7.52$\pm$0.11&0.71$\pm$0.01&9.24$\pm$0.22&0.53$\pm$0.04\\
NGC 5506            & 14h13m14.81s & -03d12m27.0s &27&2.457&90&10.41$\pm$0.10&8.05$\pm$0.08&1.96$\pm$0.08&9.06$\pm$0.18&0.65$\pm$0.07\\
NGC 5953            & 15h34m32.30s & +15d11m42.0s &28&2.168&44&9.99$\pm$0.06&7.40$\pm$0.09&2.56$\pm$0.1&9.06$\pm$0.17&0.02$\pm$0.6\\
NGC 5995            & 15h48m24.91s & -13d45m28.01s &110&2.008&42&10.87$\pm$0.08&8.98$\pm$0.07&19.19$\pm$2.3&9.38$\pm$0.25&0.34$\pm$0.05\\
NGC 6890            & 20h18m18.11s & -44d48m23.0s &35&2.201&38&9.86$\pm$0.08&6.85$\pm$0.12&2.05$\pm$0.25&9.37$\pm$0.26&0.13$\pm$0.69\\
NGC 7130            & 21h48m19.52s & -34d57m04.48s &70&2.194&34&10.49$\pm$0.09&7.74$\pm$0.05&20.93$\pm$0.05&9.17$\pm$0.22&0$\pm$0\\
NGC 7496            & 23h9m47.20s & -43d25m40.01s &24&2.525&53&9.46$\pm$0.10&7.03$\pm$0.06&1.55$\pm$0.19&8.89$\pm$0.14&0$\pm$0\\
NGC 7674            & 23h27m56.70s & +08d46m45.01s &127&2.049&27&11.11$\pm$0.10&8.00$\pm$0.07&23.58$\pm$2.83&9.30$\pm$0.22&0.58$\pm$0.6\\
TOLOLO 1238-364     & 12h40m52.90s & -36d45m22.0s &47&2.095&22&9.66$\pm$0.09&7.53$\pm$0.11&5.76$\pm$0.17&9.14$\pm$0.21&0.32$\pm$0.6\\
UGC 05101           & 09h35m51.60s & +61d21m11.45s &174&2.08&0&10.92$\pm$0.08&8.25$\pm$0.09&55.21$\pm$6.63&9.59$\pm$0.29&0.65$\pm$0.03\\

\hline
\end{tabular}
\caption{
Column description: (1) Name; (2) Right Ascension; (3) Declination; (4) Distance, in units of Mpc; (5) and (6) Logarithm of the 25 mag arcsec$^{-2}$ isophotal diameter and inclination angle in degrees, both from the Hyperleda catalog \citep{Makarov14}, respectively; (7) Logarithm of the stellar mass by G16; (8) Logarithm of the dust mass by G16; (9) SFR, in units of M$_{\odot}$ yr$^{-1}$ by G16; (10) Oxygen abundances derived through the empirical N2 relation taken from \cite{PerezMonteroContini09}; (11) the relative contribution of the AGN to the 5-40 $\mu$m band luminosity.
$^{a}$ for NGC 5256 we used the D$_{25}$ measurement from the 3RC catalog \citep{3RC}.
Since this object is a major merger, the authors did not provide any estimate for the inclination.
To provide the aperture correction factor for NGC 5256 consistently with the remaining part of the sample, we estimated the aperture correction assuming ten evenly spaced inclination angles between face-on ($i=0^{\circ}$) and edge-on ($i=90^{\circ}$) configurations. 
We then adopted a mass correction factor $f_{ap}=1.9\pm0.2$, which is the mean of the estimated values and the standard deviation as uncertainty.
}
\label{table:sample}      

\end{sidewaystable*}

%
\subsubsection{Literature single-dish data}
\label{sec:lit_sd_data}
To extend the sample coverage, we included in the analysis a set of low-J CO emission line fluxes retrieved from the literature for ten additional Seyfert 2 galaxies from G16 (references are reported in Table \ref{table:CO}).
In particular, we searched for CO(1-0) and CO(2-1) spectroscopy obtained with single-dish telescopes to avoid the filtering out of the flux due to missing short baselines, inherent to the interferometric observations\footnote{An interferometer is limited by the minimum spacing of its antennas. Two antennas cannot be placed closer than some minimum distance ($D_{min}$) and signals on spatial scales larger than some size ($\propto\lambda/D_{min}$) will be resolved out.}.
More precisely, from \cite{Papadopoulos12}, we retrieved the CO(1-0) emission line intensities of four objects (Mrk 0273, NGC 5135, NGC 5256, UGC 05101), obtained with the IRAM 30~m antenna ($\theta_{mb}\sim23$\as).
The CO(1-0) line intensities for NGC 0034 and NGC 7130 were measured by \cite{Albrecht07} with the 15~m antenna of the Swedish-ESO Submillimeter Telescope (SEST; $\theta_{mb}\sim45$\as).
In the work by \cite{Maiolino97}, the author exploited the 12~m single-dish facility of the National Radio Astronomy Observatory (NRAO) to study the molecular gas properties of a large sample of local galaxies, among which, we retrieved the CO (1-0) flux measurements for NGC 0513, NGC 3079 and NGC 5347.
Finally, \cite{Rosario18} provided the flux of the CO(2-1) transition for NGC 4388, observed with the 15~m dish of the James Clerk Maxwell Telescope (JCMT; $\theta_{mb}\sim22$\as). 
%
\subsubsection{Aperture correction for the CO flux}
\label{sec:aperture_correction}
Proprietary data from APEX as well as literature data are single-dish observations pointed at the center of the galaxy (i.e. at the optical position), with a typical field-of-view (FoV) smaller than the dimension of the optical emission from the galaxy.
To account for potential CO flux loss, we applied aperture correction to the CO line flux based on the relation between the galactic extension determined through optical observations and CO maps.
The spatial distribution of the molecular gas, traced by the CO emission, is well described by an exponentially decreasing disk, both perpendicularly to the galactic plane and in the radial direction.
The CO scale radius has been shown to be proportional to the optical size of the sources (D$_{25}$\footnote{D$_{25}$ is the major axis isophote at which the optical (band B) surface brightness falls beneath 25 mag arcsec$^{-2}$.}; e.g., \citealt{Lisenfeld11}; \citealt{Casasola17}).
Following \cite{Boselli14} we assumed: 
\begin{equation}
    S_{CO}(r,z) = S_{CO}(0) e^{-r/r_{CO}} e^{-|z|/z_{CO}},
\end{equation}
where $S_{CO}(0)$ is the central emission, $r_{CO}$ and $z_{CO}$ are the CO scale radius and height, respectively.
This method is the 3D extension of the 2D approach proposed by \cite{Lisenfeld11}, valid for low-inclination galaxies.
Here we assumed $r_{CO}/r_{25}=0.2$\footnote{r$_{25}$=D$_{25}$/2.}, following \cite{Lisenfeld11}, and  $z_{CO}/z_{25}=0.01$, as suggested by \cite{Boselli14}.
These assumptions have been tested in nearby galaxies with similar morphological classification to the Seyfert 2 galaxies (mostly spirals and S0 objects; see also \citealt{Boselli14}; Ca20).
Inclination angles and optical diameters of the Seyfert 2 galaxies are reported in Table \ref{table:sample}.
The resulting aperture correction factor is:
\begin{equation}
    f_{ap} = S_{CO, tot} ~/ ~S_{CO, mb},
\label{eq:fap}
\end{equation}
where $S_{CO, tot}$ is the total CO flux integrated over the entire galaxy, while $S_{CO, mb}$ is the scaled CO flux measured in the center of the galaxy, convolved with the main beam profile.\\
The estimated values for the aperture correction are reported in Table \ref{table:CO}.
The mean aperture correction factor is $f_{ap} = 2.8$ and standard deviation is $\sigma_{f_{ap}} = 1.5$; a large fraction of the sample (19 out of 33 targets, by $\sim$60\%) have $f_{ap}<2$, i.e. the resulting CO flux is increased by a factor lower than 2.
The only exception is NGC 4388, whose large optical dimension (D$_{25}\sim5^{\prime}$) led to $f_{ap}\sim15$.  
The CO(1-0) luminosity presented in Table \ref{table:CO} are derived from aperture corrected fluxes assuming equation \ref{eq:fap}, while the uncertainties on $L^{\prime}_{\rm CO(1-0)}$ include a contribution from the error on $f_{ap}$.

\begin{figure}[htbp]
	\includegraphics[width = 0.5\textwidth]{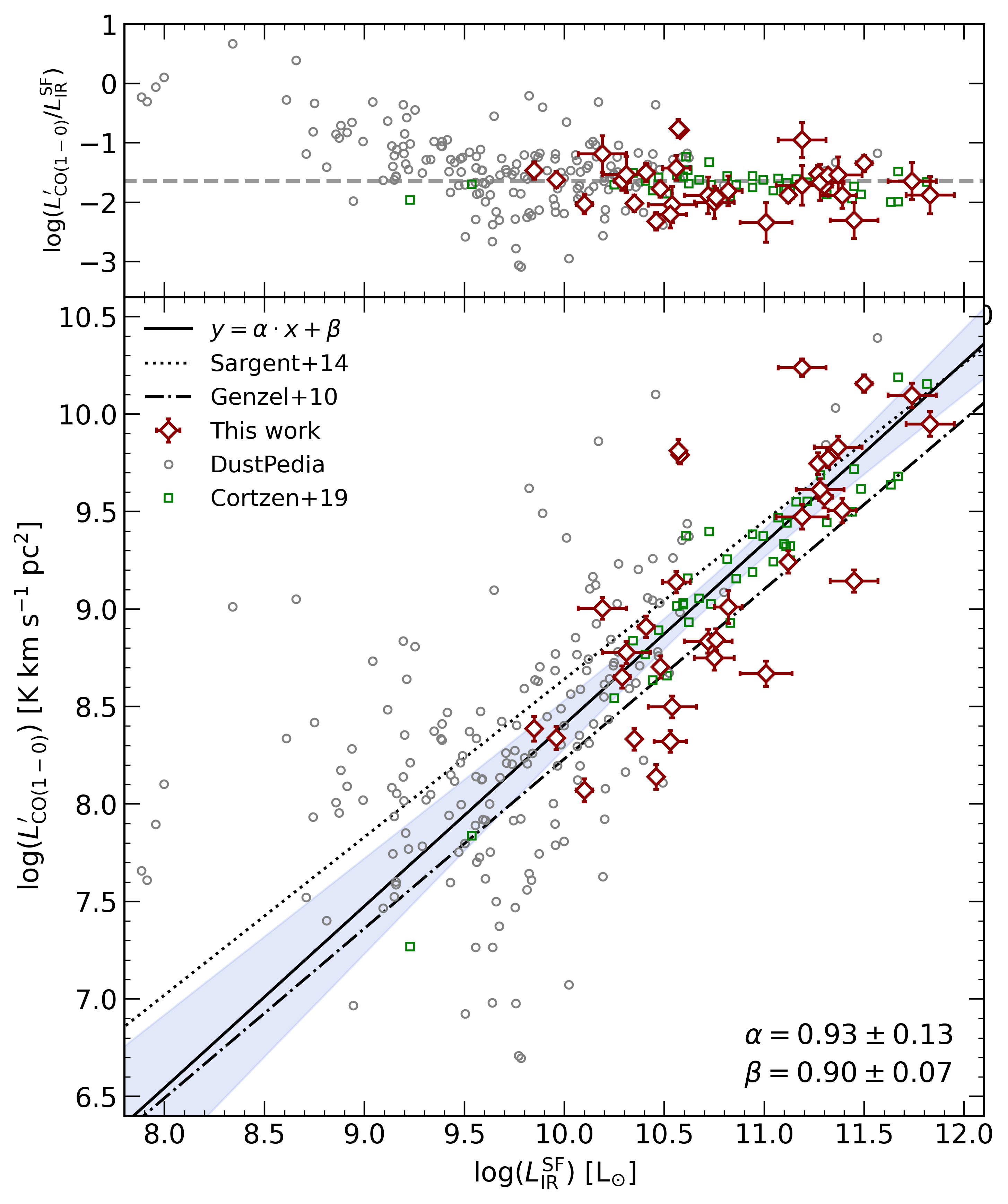}\\
	\caption{Bottom: aperture corrected CO(1-0) luminosity ($L^{\prime}_{\rm CO(1-0)}$) vs IR luminosity ($L_{\rm IR}$), for the Seyfert 2 galaxies (red diamonds).
	The best-fit parameters ($\alpha=0.93\pm0.13$, $\beta=0.90\pm0.07$) were obtained with a MCMC regression analysis, and are represented by the black solid line, while the light-blue area covers the parameters space between the 16th and 84th percentile.
	$L^{\prime}_{\rm CO(1-0)}-L_{\rm IR}$ relation for SFGs from literature are reported as dotted \citep{Sargent14} and dot-dashed \citep{Genzel10} lines.
	The control sample is shown with gray circles (DustPedia) and green squares (Co19).
	Top: $L^{\prime}_{\rm CO(1-0)}/L_{\rm IR}$ ratio as a function of $L_{\rm IR}$; the horizontal gray-dashed line is the median value.
	}
	\label{fig:Lco-Lir}
\end{figure}

%
\subsubsection{The molecular gas content}
\label{sec:Mgas}
Low-rotational transitions of the $^{12}$CO molecules -- the second most abundant molecule in the ISM \citep{YoungScoville91} -- are widely used as tracers of the cold molecular gas components in galaxies.
By measuring the total luminosity of the CO(1-0) emission line (\lco), we are able to estimate the molecular gas reservoir.  
Here, we derived the CO(1-0) luminosity by applying the correction for aperture to the CO intensities presented in Table \ref{table:CO}.
For the 24 sources with the CO(2-1) line intensity, we assumed an intensity ratio $R_{21}=I_{CO(2-1)}/I_{CO(1-0)}=0.9$ (corresponding to a flux ratio $\sim$3.6) to extrapolate the intensity of the lowest-J transition, as observed for similar objects in the local Universe (e.g. \citealt{Papadopoulos12}).
Given the CO(1-0) intensity, we calculated $L^{\prime}_{\rm CO(1-0)}$ in units of \uLco\ following \cite{Puschnig20} equation (3), which is adapted from the equation (2) from \cite{SolomonVandenBout05} with the inclusion of the aperture correction factor:
\begin{equation}
L^{\prime}_{CO}=23.5 f_{ap} \Omega I_{CO} D_{L}^2 (1+z)^{-3},
\end{equation}
where $f_{ap}$ is the aperture correction factor, $\Omega$ the solid angle of the Gaussian beam in arcsec$^{2}$, $I_{CO}$ the CO integrated line intensity in K \kms, $D_L$ the luminosity distance in Mpc, $z$ the redshift.
\\
The resulting \lco\ are reported in Table \ref{table:CO} and plotted in Fig. \ref{fig:Lco-Lir} they are plotted as a function of the IR luminosity, produced by SF ($L_{IR}^{SF,}$\footnote{Hereafter, we refer to the IR luminosity associated with SF generically, i.e. where the contribution from the AGN was ruled out, as IR luminosity.}, one of the outcomes of the SED decomposition performed by G16); further discussion of this figure will be reported
in Section \ref{sec:results}.\\
The cold molecular gas mass (M$_{\rm gas}$) is usually derived from the luminosity of the CO(1-0), by assuming a CO-to-H$_{2}$ conversion factor:
\begin{equation} 
\label{eq:Mco}
M_{\rm gas}=\alpha_{CO} L^{\prime}_{\rm CO(1-0)},
\end{equation}
as in \cite{SolomonVandenBout05}, and \cite{Bolatto13}.
Theoretical and observational studies suggested that the \aco\ factor assumes a large range of values depending on the galaxy properties (e.g., compactness, merging), the physical conditions (presence of intense radiation fields) and composition (metallicity) of the ISM (e.g., \citealt{Leroy11}; \citealt{Narayanan12}; \citealt{Papadopoulos12}; \citealt{Bolatto13}; \citealt{Sandstrom13}).
Low CO-to-H$_{2}$ conversion factors (\aco=0.3-2.5 \uaco) have been measured in local ultra-luminous infrared galaxies (U-LIRGs) and starburst galaxies, while higher values were observed in Milky Way-like objects and main-sequence (MS) galaxies (\aco=4.3\uaco; e.g., \citealt{Solomon97}; \citealt{Tacconi06}; \citealt{Daddi10b}; \citealt{Magdis11}; \citealt{Bolatto13}; \citealt{Magdis13}; \citealt{Casey14}).
Indeed, relatively low \aco values ($\sim$1.1 \uaco) were used to estimate \mh\ in the central metal-rich regions of local active and inactive galaxies, with properties (e.g., M$_{\star}$, SFR) similar to those of the sample of Seyfert 2 galaxies (e.g., \citealt{Pozzi17}; \citealt{Rosario18}).\\
Given the diverse nature of the Seyfert 2 galaxies presented in this work, which include several LIRGs, we adopted the prescription by \cite{Narayanan12} to determine the proper \aco\ factor for each object, i.e.:
\begin{equation} 
\label{eq:alpha_co}
\alpha_{CO} = \frac{min(6.3;10.7\times I_{CO}^{-0.32})}{Z^{0.65}},
\end{equation}
where Z is the metallicity, derived from the O/H abundance in proportion to the solar abundance (see Section \ref{sec:metallicity}), and $I_{CO}$ is the CO brightness intensity $I_{CO}$.
Equation \ref{eq:alpha_co} is based on a semi-analytic relation by \cite{Narayanan12} where the \aco\ conversion factor depends on both the metallicity and the CO intensity ($I_{CO}$).
The dependence on the CO brightness makes \aco\ sensitive to varying environmental properties, such as diverse density and temperature of the molecular gas.
This is even more true in ULIRGs, where starbursting regions would make the molecular gas denser and hotter, thus resulting in brighter CO emission (and thus lower \aco; for further details, see also the Appendix in \citealt{Puschnig20}).
We obtained \aco\ values in the range \aco=0.6-5.2 \uaco, with $\sim$90\% of the sample (29/33 objects) having a \aco\ between 1.1 and 4.3 \uaco, i.e., the \aco\ usually adopted for AGN and Milky-way like galaxies, respectively.
Only 4 targets show \aco\ outside this range: NGC 3079 and NGC 5135 with \aco$<1.1$\uaco (i.e., the galaxies with the higher oxygen abundance in the sample; see Table \ref{table:sample}), IC 5063 and Mrk 0463 with \aco$>4.3$\uaco (i.e., the galaxies with the lower oxygen abundance in the sample).
These two couples of objects show Z measurements at the higher and lower tails of the gas-phase metallicity distribution, respectively, and there are no evidences suggesting for the 4 galaxies a nature different from that of the rest of the sample.
In Table \ref{table:CO}, for each galaxy we present the best estimate for the adopted \aco\ and \mh\, derived by using equation \ref{eq:Mco} and including the helium contribution, along with the corresponding uncertainties.
In Section \ref{sec:results}, we graphically show how the estimate for the \mh\ would change by assuming \aco\ values in the range \aco=1.1-4.3 \uaco\ (i.e. the CO-to-H$_2$ conversion factors usually adopted for local AGN and Milky-way like objects, respectively), and how the \aco\ assumption affects the results.

%
\begin{sidewaystable*}

\centering          
\begin{tabular}{ l l l l l l l l l l l}     
\hline     
Name & Spectral line & $I_{CO}$[2-1] & $S_{CO}$[2-1] &  rms & $W_{CO}$ & $\log(L^{\prime}_{CO(1-0})$ & \aco\ & $\log(M_{H_2})$ &$f_{ap}$& Ref.\\ 
     &               &   K \kms      &  Jy \kms      &  mK  &   \kms  & $\log(L_{\odot})$ & \uaco\ & $\log(M_{\odot})$&&  \\ 
(1)  &  (2)          &   (3)         &  (4)          &  (5)  &  (6)  & (7) &(8)  &(9)&  (10) & (11)  \\ 
\hline                   
CGCG 381-051       &  CO(2-1)    &3.4$\pm$0.3&127$\pm$15&0.54&187$\pm$3&9.24$\pm$0.06&4.30$\pm$1.01&9.88$\pm$0.10&1.31$\pm$0.05&     (a)    \\
ESO 033-G002       &   CO(2-1)   &0.67$\pm$0.07&25$\pm$3&0.32&161$\pm$6&8.14$\pm$0.06&2.88$\pm$1.00&8.60$\pm$0.15&1.54$\pm$0.07&     (a)    \\
IC 5063            &  CO(2-1)    &3$\pm$0.29&111$\pm$13.96&0.63&385$\pm$5&8.84$\pm$0.07&4.57$\pm$1.05&9.50$\pm$0.10&4.40$\pm$0.13&     (a)    \\
IRAS F01475-0740    &   CO(2-1) &2.23$\pm$0.18&82.65$\pm$9.49&0.58&87.8$\pm$1.1&8.49$\pm$0.06&4.17$\pm$1.06&9.11$\pm$0.11&1.11$\pm$0.01&     (a)    \\
IRAS F04385-0828    &   CO(2-1) &1.77$\pm$0.14&65.35$\pm$7.49&0.25&142$\pm$1.6&8.32$\pm$0.06&2.69$\pm$0.99&8.75$\pm$0.16&1.27$\pm$0.07&     (a)    \\
IRAS F15480-0344    &   CO(2-1) &1.1$\pm$0.12&40.7$\pm$5.41&0.5&150$\pm$5&8.67$\pm$0.08&4.07$\pm$1.13&9.28$\pm$0.12&1.13$\pm$0.04&     (a)    \\
MCG-03-34-064       &   CO(2-1) &2.56$\pm$0.26&94.66$\pm$12.34&0.75&334$\pm$11&8.75$\pm$0.06&1.55$\pm$0.75&8.94$\pm$0.21&1.97$\pm$0.15&     (a)    \\
MCG-03-58-007       &   CO(2-1) &6.57$\pm$0.54&243.14$\pm$28.16&0.73&319$\pm$4&9.58$\pm$0.06&1.95$\pm$0.81&9.87$\pm$0.18&1.41$\pm$0.02&     (a)    \\
MCG+00-29-023       &   CO(2-1) &15.45$\pm$1.21&571.82$\pm$64.42&0.5&361.2$\pm$1.1&9.75$\pm$0.05&2.51$\pm$0.69&10.15$\pm$0.12&1.42$\pm$0.07&     (a)    \\
Mrk 0273            &   CO(1-0)     &  &&  & &9.95$\pm$0.06&1.45$\pm$0.60&10.11$\pm$0.18&1.38$\pm$0.03& (b)    \\
Mrk 0463            &   CO(2-1)  &1.74$\pm$0.18&64.39$\pm$8.43&0.53&228$\pm$7&9.47$\pm$0.06&5.25$\pm$0.97&10.19$\pm$0.08&1.62$\pm$0.18&     (a)    \\
Mrk 0897            &   CO(2-1) &3.75$\pm$0.31&138.87$\pm$16.2&0.6&216$\pm$3&9.14$\pm$0.07&1.86$\pm$0.82&9.41$\pm$0.19&1.31$\pm$0.01&     (a)    \\
NGC 0034            &   CO(1-0)     &&&&&9.51$\pm$0.06&1.62$\pm$0.75&9.72$\pm$0.20&1.29$\pm$0.04& (c)    \\
NGC 0424            &   CO(2-1) &0.9$\pm$0.08&33.4$\pm$4.02&0.17&258$\pm$6&8.07$\pm$0.06&3.39$\pm$1.09&8.6$\pm$0.14&2.31$\pm$0.04&     (a)    \\
NGC 0513            &   CO(1-0)     &&&&&9.01$\pm$0.08&3.16$\pm$1.24&9.51$\pm$0.17&1.06$\pm$0.05& (d)    \\
NGC 1125            &   CO(2-1) &2.1$\pm$0.17&77.84$\pm$8.87&0.21&228$\pm$2&8.33$\pm$0.06&3.16$\pm$1.09&8.83$\pm$0.15&2.12$\pm$0.09&     (a)    \\
NGC 1320            &   CO(2-1) &2.61$\pm$0.23&96.58$\pm$11.69&0.43&338$\pm$8&8.34$\pm$0.10&3.31$\pm$1.07&8.86$\pm$0.14&2.60$\pm$0.20&     (a)    \\
NGC 2992            &   CO(2-1) &12.99$\pm$1.02&480.52$\pm$54.25&0.52&492$\pm$3&9.14$\pm$0.06&2.04$\pm$0.71&9.45$\pm$0.15&4.41$\pm$0.01&    (a)    \\
NGC 3079            &   CO(1-0)     &&&&&9.79$\pm$0.05&0.59$\pm$0.33&9.56$\pm$0.24&7.18$\pm$0.3& (d)    \\
NGC 4388            &   CO(2-1)     &  &&  &  &9.81$\pm$0.06&3.02$\pm$0.56&10.29$\pm$0.08&15.87$\pm$0.02&  (e)    \\
NGC 4602            &   CO(2-1)  &8.96$\pm$0.74&331.56$\pm$38.28&1.38&148.1$\pm$1.9&8.70$\pm$0.06&3.39$\pm$0.86&9.23$\pm$0.11&1.92$\pm$0.03& (a)    \\
NGC 5135            &   CO(1-0)     &  &&  &  &10.24$\pm$0.04&0.69$\pm$0.33&10.08$\pm$0.21&5.10$\pm$0.06&  (b)    \\
NGC 5256            &   CO(1-0)     &  &&  &   &10.16$\pm$0.04&1.70$\pm$0.51&10.39$\pm$0.13&2.20$\pm$0.20& (b)    \\
NGC 5347            &   CO(1-0)     &&&&&8.39$\pm$0.06&2.69$\pm$0.93&8.82$\pm$0.15&1.41$\pm$0.11& (d)    \\
NGC 5506            &   CO(2-1)  &6.68$\pm$0.54&246.99$\pm$28.37&0.7&327$\pm$4&8.65$\pm$0.06&3.24$\pm$0.89&9.16$\pm$0.12&4.31$\pm$0.01&     (a)    \\
NGC 5953            &   CO(2-1)  &20.88$\pm$1.63&772.68$\pm$87.02&0.72&204.3$\pm$0.7&8.91$\pm$0.10&2.24$\pm$0.62&9.26$\pm$0.12&2.20$\pm$0.04&     (a)    \\
NGC 5995            &   CO(2-1)  &9.82$\pm$0.78&363.27$\pm$41.13&0.5&420$\pm$3&9.61$\pm$0.06&1.78$\pm$0.74&9.86$\pm$0.18&1.61$\pm$0.07&     (a)    \\
NGC 6890            &   CO(2-1)  &9.45$\pm$0.78&349.82$\pm$40.38&1.21&237$\pm$3&8.78$\pm$0.06&1.78$\pm$0.70&9.03$\pm$0.17&2.38$\pm$0.03&     (a)    \\
NGC 7130            &   CO(1-0)     &  &&  &   &9.77$\pm$0.04&2.19$\pm$0.71&10.11$\pm$0.14&1.58$\pm$0.08& (c)    \\
NGC 7496            &   CO(2-1) &13.52$\pm$1.06&500.22$\pm$56.42&1.23&89.4$\pm$0.4&9.00$\pm$0.06&3.31$\pm$0.76&9.52$\pm$0.10&6.10$\pm$0.30&     (a)    \\
NGC 7674            &   CO(2-1)  &11.44$\pm$0.9&423.34$\pm$47.9&0.79&194.8$\pm$1.2&9.83$\pm$0.06&1.86$\pm$0.64&10.10$\pm$0.15&1.73$\pm$0.18&     (a)    \\
TOLOLO 1238-364     &   CO(2-1) &7.58$\pm$0.61&280.62$\pm$32.08&0.93&151.5$\pm$1.9&8.84$\pm$0.8&2.75$\pm$0.95&9.28$\pm$0.15&1.90$\pm$0.03&     (a)    \\
UGC 05101           &   CO(1-0)     &  &&  &   &10.10$\pm$0.06&1.15$\pm$0.56&10.16$\pm$0.21&2.20$\pm$0.30& (b)    \\

\hline                  
\end{tabular}
\caption{Column description: (1) Source name; (2) CO transition considered; (3) CO integrated line intensity in units of \Kkms from the new APEX observations; (4) CO flux in units of \jykms; (5) RMS of the CO line in units of mK; (6) CO line width in units of \kms; (7) Logarithm of the aperture-corrected $L^{\prime}_{CO(1-0}$ in \uLco; (8) Logarithm of the aperture-corrected molecular gas mass ($M_{H_2}$) in units of \ms; (9) Adopted aperture correction factor with the associated error (see Sect. \ref{sec:aperture_correction}).
(10) References to the CO spectroscopy: (a) This paper; (b) \cite{Papadopoulos12};  (c) \cite{Albrecht07}; (d) \cite{Maiolino97}; (e) \cite{Rosario18}.} 
\label{table:CO}      
\end{sidewaystable*}


%
%
\subsection{Optical emission lines}
\label{sec:metallicity}
The \aco\ proposed by \cite{Narayanan12} requires the determination of the gas-phase metallicity in units of solar metallicity.
Here, we assume that the oxygen abundance is a good tracer of the total gas-phase metallicity.
Among the diverse calibrations present in the literature to derive the oxygen abundance, we adopted the empirical calibration by \cite{PettiniPagel04}, based on N2 index:
\begin{equation}
\label{eq:OH}
    12 + log(O/H) = 9.37 + 2.03\times N2 + 1.26\times N2^2 + 0.32 \times N2^3.
\end{equation}
where $N2 = \log([NII] \lambda6583\r{A} /H\alpha)$.
We opted for equation \ref{eq:OH} because the N2 index only requires the flux ratio of $[NII] \lambda6583\r{A}$ and $H\alpha$ emission lines, which is available in the literature for the entire sample.
Optical line flux measurements for 31 out of the 33 Seyfert 2 galaxies are from ancillary UV/optical spectra analyzed by \cite{Malkan17}, while for the remaining two objects (IRAS F04385-0828, NGC 4602) we collected the optical line ratios from the optical spectra obtained with the South African Large Telescope (SALT) by Feltre et al. (in prep.).
Errors on the optical emission lines include a contribution from the calibration uncertainty, which we conservatively assumed to be of 30\% (see \citealt{Malkan17} for further details).
In Table \ref{table:sample}, we reported the oxygen abundance with the associated uncertainty.

%
%
\subsection{Decomposed SED}
\label{sec:sed_decomposition}
\begin{figure}[htbp]
	\includegraphics[width = 0.5\textwidth, keepaspectratio=True]{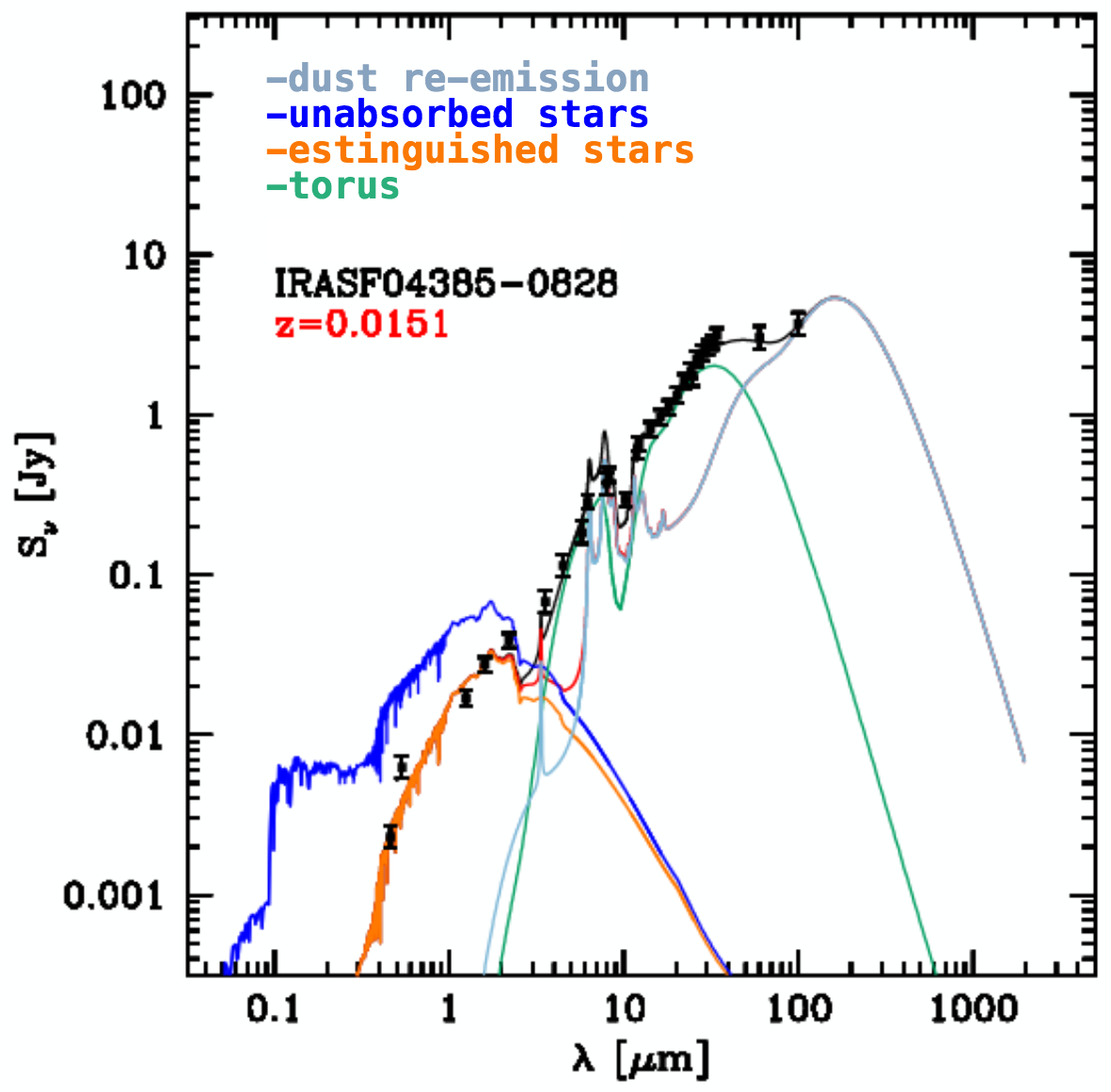}\\
	\caption{Example of decomposed SED (from G16): unabsorbed (blue line) stellar component is scaled down due to the dust-absorption to match observations (black dots); the resulting absorbed stellar emission is represented in orange. In the MIR band, the dusty torus component is shown with the green-dashed line, while the dust re-emission in the FIR is in gray.
	}
	\label{fig:SED}
\end{figure}

The sources in the sample of Seyfert 2 galaxies benefit from the detailed SED decomposition analysis performed by G16, which provides a complete description of each source in terms of the different components (e.g., stars, dust, AGN) and the ongoing processes (e.g., SF, nuclear accretion). 
Here we briefly summarize the approach adopted in G16, where the SED fitting procedure and the processing of the selection of the archival data are illustrated with great details.
The SED-fitting code adopted is \texttt{SED3FIT} (\citealt{Berta13}\footnote{http://steatreb.altervista.org/alterpages/sed3fit.html}) that reproduces simultaneously the stellar emission and the reprocessed emission from the dust -- heated by both stars and the AGN.
The code relies on a collection of libraries, in particular the library by \cite{BruzualCharlot03} for the stellar contribution, the one by \cite{daCunha08} for the IR dust-emission, and the library of AGN tori by \cite{Fritz06}, updated by \cite{Feltre12}.
An ancillary compilation of photometric data, from the UV to the FIR wavelengths, were collected from NED.
Furthermore, to properly constrain and disentangle the AGN contribution, with the dusty torus contributing the most in the MIR, archival \sirs\ data were included (see G16 for further details).
The sources in the sample being optically classified as Seyfert 2 galaxies, i.e. narrow-line/obscured AGN, the contribution from stars dominates over the AGN in the optical/UV band, making it easier to differentiate between the contributions from AGN and host galaxy in the global outcome of the source.
This led to a more reliable characterization of the source in terms of the stellar content and the SF activity with respect to sources with an unobscured AGN.\\ 
An example of decomposed SED for one of the objects (IRAS F04385-0828) is presented in Fig. \ref{fig:SED}.
In Table 3 of G16, the authors reported the results of the SED decomposition for their entire sample, among which the 33 Seyfert 2 galaxies can be found.
For our purposes, we retrieved the main host-galaxy properties, namely the SFR (obtained through the \citealt{Kennicutt98} relation), stellar and total dust masses (measured as prescribed by \citealt{daCunha08}), IR luminosity (integrated over the 8-1000$\mu$m spectral range), as well as the relative contribution of the AGN (f$_{AGN}$) to the global outcome of the source in the 5-40 $\mu$m band (f$_{AGN}$), that we present in Table \ref{table:sample}. 

%
%
\subsection{MIR features}
\label{sec:PAH}
The \sirs\ MIR spectra of the central region of local sources offer a wealth of spectral features which are fundamental diagnostics of the SF vs. AGN interplay.
This spectral range is characterized by the concurrent contributions from the thermal continuum emission from the dust associated with SF, as well as spectral features and lines arising from the different gas component (molecular, atomic and ionized) and the dust reprocessed emission from the AGN, as in the case of the Seyfert 2 galaxies.
Many of these spectral features have been widely used to determine the impinging mechanism responsible for the observed emission, among which the PAH features, which are associated with SF activity, but can be affected by the presence of strong radiation fields from the AGN.\\
We collected measurements of MIR features from the work by HC11 for 32 out of the 33 Seyfert 2 galaxies.
In HC11, the authors analyzed \sirs\ spectra of 739 sources, both active and SFGs with redshift up to $\sim3.7$, gathered from many observational campaigns (see Table 1 in HC11 for more details and references).
HC11 provided the MIR measurements such as the PAH main properties, the strength of the silicates in emission or absorption around 9.7 $\mu$m, rest-frame monochromatic luminosities or colors, used as diagnostics to classify the sources in terms of their MIR properties.
In the case of local objects from HC11 -- including the 32 in common with the present sample of Seyfert 2 galaxies -- the \sirs\ spectra sample the emission from the nuclear region, where the contribution from the AGN is more relevant.
The actual extraction area depends on the slit mode and the distance of the source\footnote{See also \url{https://irsa.ipac.caltech.edu/data/SPITZER/docs/files/spitzer/irs_pocketguide.pdf} for further details on the \sirs\ specifications.}, but the central kpc-scale region was sampled even in the most nearby objects in the sample.\\
Since a significant fraction of the \sirs\ spectra included in the work by HC11 had a low signal-to-noise ratio, a proper modelling of the MIR features was difficult.
For this reason, the authors defined a homogeneous and concise method providing solid estimates for each source, by combining the linear interpolation of the continuum with the integration of the emission from the features.
In particular, in the case of the PAH features we are interested in, the authors selected two narrow, continuum bands at both sides of each feature, performed a linear interpolation to estimate the continuum underlying the feature, then subtracted it from the spectrum.
The residuals were integrated in a band centered at the expected wavelength of the peak of the PAH feature to obtain the integrated PAH flux.
The uncertainties on the PAH intensities and the continuum were estimated by performing Monte Carlo simulations.
The authors also provided the equivalent widths (EWs) of the PAH features by dividing the integrated PAH flux by the interpolated continuum at the center of the feature.
For further details about the procedure, we refer to Section 4 in HC11.\\
We collected the luminosity and EW for the 6.2 and 11.3 $\mu$m PAH features for 28 and 32 out of the 33 Seyfert 2 galaxies, respectively.
%
%
%
\begin{figure*}[pt]
	\includegraphics[width = \textwidth, keepaspectratio=True]{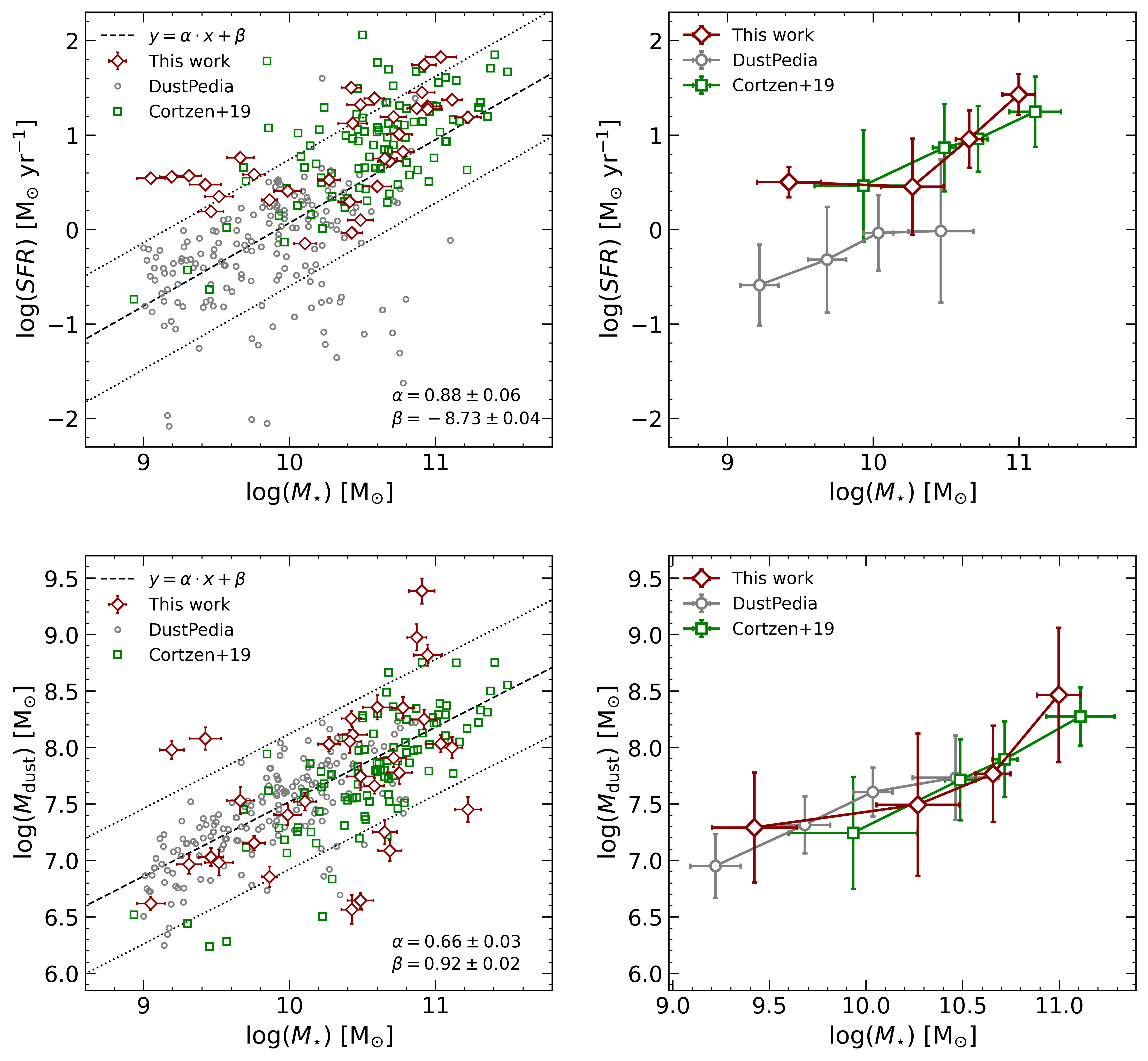}\\
	\caption{Left panels: SFR (top row) and dust mass (bottom row) as a function of the stellar mass ($M_{\star}$) for Seyfert 2 galaxies (red diamonds) and the control sample (gray circles for DustPedia galaxies and green squares for 5MUSES sources).
	For pure illustrative purposes, the best-fit trend (black-dashed line) for the control sample of inactive galaxies is shown, while the two black-dotted lines are the same trend shifted by a factor of 0.5~dex in either direction.
	The best-fit slope ($\alpha$) and normalization ($\beta$) are reported in each panel.
	Right panels: same SFR-$M_{\star}$ and M$_{dust}$-$M_{\star}$ planes as in the left panels, but with binned data to underline the average trend of active and inactive galaxies.
	The bins were chosen to include the same number of sources.
	The mean value and the error bars (standard deviations) in each bin were determined with a bootstrap procedure using 10000 iterations.
}
	\label{fig:comparison_cs}
\end{figure*}
%

\begin{table*}[ht]
\centering          
\begin{tabular}[0.5\textwidth]{l l l l}     
\hline     
                    & Seyfert 2     & DustPedia     & 5MUSES  \\ 
\hline                  
Number              &       33       &   169              &   95              \\ 
$D$ [Mpc]          & 15-224    &  0.3-38.3        &    109-2090               \\ 
SFR [\usfr]         & 0.7 - 66.8  &  0.008 - 39.8  &   0.18 - 114.5  \\ 
$\log(M_{\star}/M_{\odot})$    & 10.3$\pm$0.6  &  9.9$\pm$0.5  &  10.6$\pm$0.5  \\ 
$\log(M_{dust}/M_{\odot})$    & 7.7$\pm$0.7    &  7.0$\pm$0.4  &  7.8$\pm$0.5     \\ 
\hline                  
\end{tabular}
\caption{Properties for the studied AGN sample and the control sample by Ca20 (DustPedia) and Co19 (5MUSES), after removing AGN-dominated objects.
From top to bottom, rows contain the size of sample; the intervals of distances (in units of Mpc) and SFR (in units of \usfr); the mean and standard deviation of the logarithm of the stellar and dust masses (both in units of $M_{\odot}$).}
\label{table:control_samples}      
\end{table*}

\section{Control sample}
\label{sec:control_sample}
To assess the effect of AGN on the properties of the host galaxy (e.g., in terms of either the molecular gas content or the SF activity), we need to compare the AGN sample with local star-forming galaxies that do not harbor an active nucleus.
Among the plethora of samples of local objects that have been studied in the literature, we focused on those samples that benefit from a complete characterization of the sources in terms of their molecular gas, dust and stellar content, as well as the SF activity.\\
The sample of local objects in the DustPedia\footnote{\url{http://dustpedia.astro.noa.gr/}} project is ideal to this purpose, given the multi-wavelength imaging and photometry database of the 875 nearby galaxies of this project (\citealt{Davies17}; \citealt{Clark18}).
The DustPedia sample consists of all the galaxies observed by \emph{Herschel}, with optical diameter $>1^{\prime}$, recessional velocity $>3000$ \kms\ and with a \emph{WISE} 3.4 $\mu$m detection with a minimum S/N of 5 \citep{Davies17}.
Each galaxy benefits from the results of a CIGALE\footnote{\url{https://cigale.lam.fr/}} SED fitting decomposition, providing the description of the sources in terms of ongoing SF and stellar content \citep{Nersesian19}.
In particular, we selected the control sample from a recent work of the DustPedia collaboration \citep[][hereafter, Ca20]{Casasola20}, focused on the molecular gas properties of a sub-sample of 255 spirals.
Using single-dish archival observations, the authors derived the molecular gas masses from aperture-corrected low-J CO spectroscopy, with a procedure similar to the one described in Section \ref{sec:Mgas} and assuming \aco=3.2 \uaco\, which is a value suited for inactive galaxies (see Ca20 for further details).
To compare the control sample with the sample of Seyfert 2 galaxies, we corrected the \mh\ from Ca20 for \aco=4.3 \uaco\, to account for the He contribution to the mass.\\
To build a control sample of inactive galaxies which matches the host-galaxy properties of the Seyfert 2 galaxies, in particular in terms of stellar mass and SFR, we excluded the dwarf galaxies by removing the objects with $M_{\star}<10^9$ \ms\ (25 objects).
Nevertheless, the remaining 230 SFGs from the sample by Ca20 still do not perfectly match the AGN sample in terms of SFR and stellar masses, as shown in Table \ref{table:control_samples}.\\
To pair the high-SFR and stellar mass tails of the sample of Seyfert 2 galaxies, we included the sample by Co19, where the authors investigated the use of MIR features, in particular the PAH features, as tracers of the molecular gas content in local and intermediate-redshift galaxies.
The target sample presented by Co19 consists of 283 MIR selected objects, drawn from the 5 mJy Unbiased \emph{Spitzer} Extragalactic Survey (5MUSES; \citealt{Wu10}) upon the availability of low-J CO spectroscopy, MIR PAH features detection, and IR photometry.
We further selected those objects (144/283) with estimates of stellar mass, SFR, molecular gas, dust and measurement of the 6.2 $\mu$m PAH feature.
They chose the PAH feature at 6.2 $\mu$m as it is less affected by the contribution from the warm dust, which is stronger at longer wavelengths.
Stellar and dust parameters were obtained through a SED decomposition \citep{Shi11}, while the SFR is derived assuming the $L_{IR}$-SFR relation by \cite{Kennicutt98}. 
Co19 also provided aperture-corrected \lco\ measurements for 33 (out of 144) 5MUSES objects, derived from CO(1-0) spectroscopy with the IRAM 30~m antenna.
The measurements of the PAH features were obtained using the \textsc{PAHFIT} code (see \citealt{Magdis13}).
Co19 provided the molecular gas masses derived from the dust content, assuming a well calibrated metallicity-dependent gas-to-dust ratio (GDR)\footnote{$\log(GDR)=(10.54\pm1.0)-(0.99\pm0.12)\times(12+\log(O/H))$ from \cite{Magdis12}.}. 
This way, the author avoided the dependency on the \aco\ conversion factor, although in this case the molecular gas masses are affected by the uncertainties related to the assumed GDR.
Nevertheless, we tested the consistency of the method to measure the \mh\ adopted by Co19 with the procedure presented in Section \ref{sec:Mgas} and in Ca20 for the objects from Co19 with \lco\ measurements.
Then, by assuming a \aco=4.3 \uaco in equation \ref{eq:Mco}, we derived the molecular gas mass and compared these values with those reported by Co19, i.e. values obtained assuming a GDR.
We checked that the two methods provide consistent estimates (within 1$\sigma$) for \mh\ but, given the larger statistics, we used the \mh\ derived from the dust content.
We refer to Co19 for the details on the measurements of the galaxy properties.
Among the 144 galaxies with a full set of PAH measurements, we conservatively excluded the sources with an EW of the 6.2 $\mu$m PAH feature smaller than 0.4 $\mu$m, which is usually adopted as an indicator for AGN or composite objects (i.e. where AGN and SF coexist; e.g., \citealt{Spoon07}; \citealt{Magdis13}; Co19), since PAH emission EW have been observed decreasing with increasing AGN activity \citep{Tommasin10}.
The remaining 95 objects are putative SFGs.\\
Since we were interested in collecting SFGs free from any AGN contamination, we further checked the potential presence of nuclear activity by cross-matching the SFGs from the DustPedia (230 objects) and 5MUSES (95 objects) samples with the most recent catalogs of X-ray observations: 4XMM-DR9\footnote{\url{http://xmmssc.irap.omp.eu/Catalogue/4XMM-DR9/4XMM_DR9.html}} \citep{Webb20} and the \emph{Swift}-BAT 105-Month Hard X-ray Survey\footnote{\url{https://swift.gsfc.nasa.gov/results/bs105mon/}} \citep{Oh18}.
96 out of 230 DustPedia sources (42\,\%) have a counterpart within 30 arcsec in the 4XMM DR9 catalog, which potentially host a relatively luminous AGN (i.e. with an observed X-ray luminosity of $L_{0.2-12 keV}>10^{41} erg s^{-1}$.
Given the low threshold adopted, we expect to point out also intrinsically weak or extremely obscured objects (e.g., \citealt{Salvestrini20}), in addition to canonical AGN.
We used the hardness ratio ($HR$) as a selection criteria to infer the presence of nuclear activity.
It is calculated as $HR=(H-S)/(H+S)$, i.e. the normalized difference of the fluxes in the hard 2-12 keV (H) and soft 0.2-2 keV (S) energy bands (for further details on the flux estimates, we refer to \citealt{Webb20} and \citep{Oh18}).
The signature of nuclear activity in low and intermediate-redshift AGN is the peak of the X-ray emission in the hard band ($E>2$ keV), i.e. resulting in positive $HR$, while the diffuse emission associated with the host-galaxy SF peaks in the soft band, which means negative $HR$.
60 potential AGN with $HR>0$ were excluded in the end.
We also searched for objects from the DustPedia sample in the \emph{Swift}-BAT catalog.
Since the emission in the hard X-ray band ($E>10$ keV) is almost exclusively associated with nuclear activity, we further excluded one source which was detected in the 14-195 keV band within the \emph{Swift}-BAT monitoring.
In the end, we retrieved the properties for the 169 local inactive galaxies from the official DustPedia web-page, namely the molecular gas and dust masses, SFR, and stellar mass.\\
As previously done for the DustPedia objects, we further check for the evidence of nuclear activity, by searching for detection in the X-rays for the 5MUSES galaxies. 
Then, we cross-matched the 95 5MUSES objects with the 4XMM-DR9 and the \emph{Swift}-BAT 105-Month Hard X-ray Survey.
We found that none of the 95 5MUSE was detected within the \emph{Swift}-BAT catalog, while 41 were detected with XMM-\emph{Newton}, but all of them showed $HR<0$.
Since there were no common sources between the two sets of objects, we set as control sample for our study the 95 objects from Co19, combined with the 169 SFGs from Ca20.
From here to the remaining part of this work, we refer as the control sample to the 264 SFGs presented in this section.
A brief summary of the main properties of the control sample and the Seyfert 2 galaxies are shown in Table \ref{table:control_samples}.
We assumed the appropriate conversion factor to correct \mstar\ and SFR measurements for the initial mass function by \cite{Chabrier03}, that is the one assumed by G16.
%
\begin{figure*}[pt]
	\includegraphics[width = \textwidth, keepaspectratio=True]{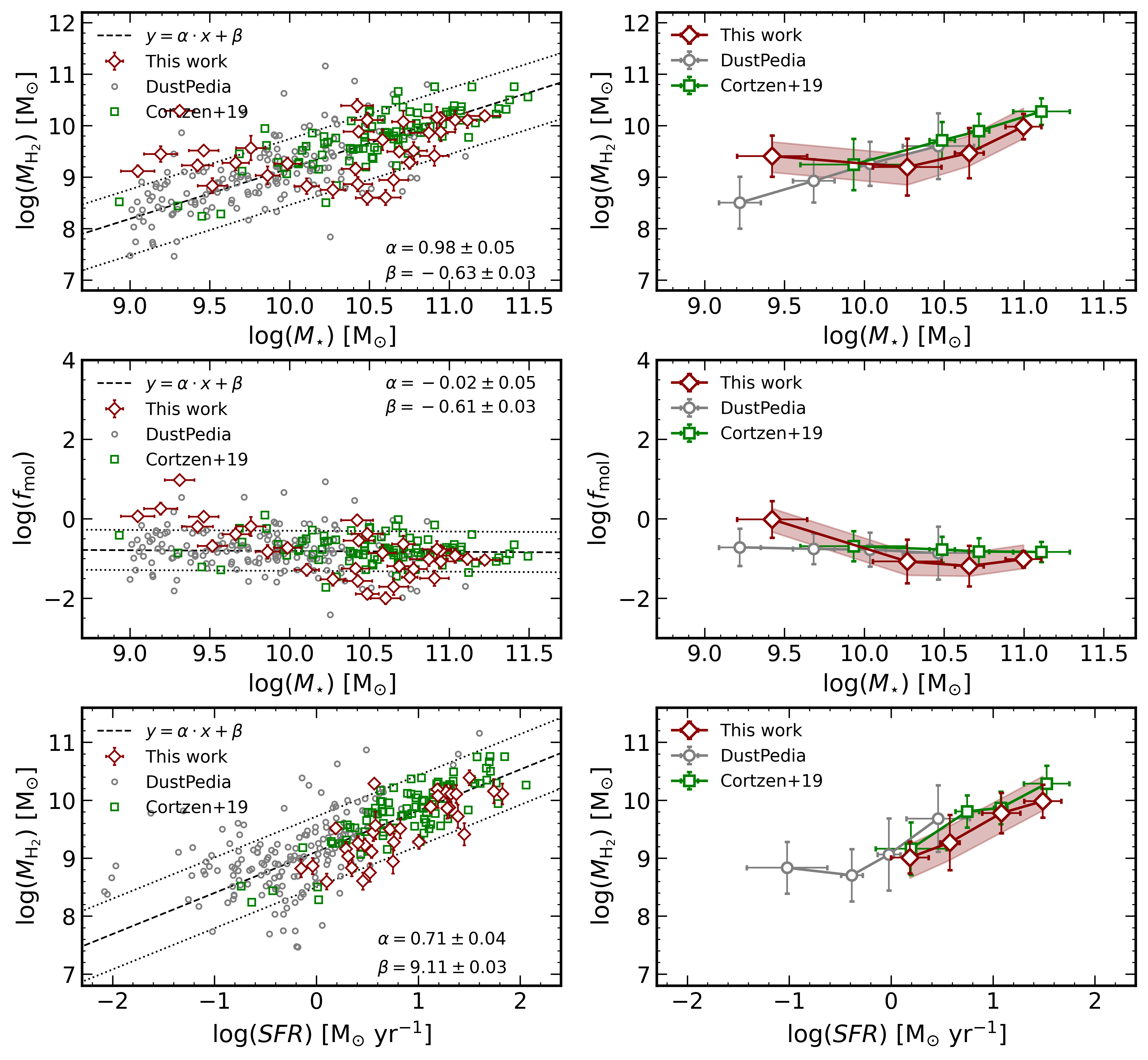}\\
	\caption{Left panels: scaling relations of the molecular gas mass ($M_{H_2}$) as a function of the host-galaxy stellar mass ($M_{\star}$; top row) and SFR (bottom row); in the central panel, we show the molecular gas fraction ($f_{mol}=M_{H_2}/M_{\star}$) as a function of $M_{\star}$.
	Seyfert 2 galaxies are shown as red diamonds, while the control sample are represented by gray circles (DustPedia) and green squares (5MUSES).
	For pure illustrative purposes, the best-fit trend (black-dashed line) for the control sample of inactive galaxies is shown, while the two black-dotted lines are the same trend shifted by a factor of 0.5~dex in either direction.
	The best-fit slope ($\alpha$) and normalization ($\beta$) are reported in each panel.
	Right panels: same scaling relations as in the left panels, but with binned data to underline the average trend of active and inactive galaxies.
	The red-shaded regions represent the range of \mh\ values of AGN obtained assuming an \aco\ in the range between the Milky-way like value (\aco=4.3\uaco) and \aco=1.1\uaco, typically used for the central region of local AGN (see the main text for further details).
	The bins were chosen to include the same number of sources.
	The mean value and error bars (standard deviations) in each bin were determined with a bootstrap procedure using 10000 iterations.
}
	\label{fig:scal_rel}
\end{figure*}

%
%
\section{Results and discussion}
\label{sec:results}
In this Section, we discuss the relations between the properties of the sample of Seyfert 2 galaxies and the control sample of SFGs.
In particular, we focus on the molecular gas content, traced by the CO emission, and physical properties such as \lir, SF, PAH features emission, as well as stars and dust content.
To compare the population of AGN considered in this study with the control sample of SFGs, we used the standard Kolmogorov-Smirnov (KS) test for two samples.
We assumed as a threshold for the p-value $p=0.05$, above which we cannot reject the null-hypothesis that the samples are drawn from the same distribution.
It is worth noticing that the results of the KS tests used to statistically compare the samples (AGN and SFGs) may be affected if the uncertainties on the galaxy properties are not properly taken into account.
To tackle this issue we simulated 1000 copies of the samples of AGN and SFGs, with each of their properties (e.g., \mstar, \mdust, SFR) randomly drawn from a normal distribution centered on the best estimate (see Table \ref{table:sample} and \ref{table:CO} for the sample of AGN, Ca20 and Co19 for the SFGs), with the relative uncertainty as standard deviation.
Thus, we were able to limit the impact, if present, of poorly constrained measurements in shaping the one-dimensional distribution of the physical properties of both AGN and SFGs samples.
Moreover, simulations may also reduce the impact of the lack of homogeneity between the selection criteria adopted to build the AGN control sample.
It is worth mentioning that KS test is most sensitive when the empirical distribution functions differ in a global fashion near the center of the distribution, while it less sensitive in the case of the difference arise in the wings of the distribution curves.
An alternative to the KS test is represented by the Anderson-Darling (AD) test \citep{AndersonDarling}, which is more sensitive to the distribution wings, but it is not recommended for small samples (as in the case of the Seyfert 2 galaxy sample).
Since both the KS and AD tests have limitations, we run both test to highlight any difference between the AGN and SFGs simulated samples.
For the sake of simplicity, in the following sections we only present the median of the p-values obtained by the KS tests, since we do not find conflicting outcomes from KS and AD tests. 
The distribution of the KS test p-values on the simulated copies of the AGN and SFGS samples are shown as histograms in Appendix \ref{app:KS-test}.
%
%
\subsection{AGN and control sample properties}
\label{sec:compar_cs}
At first, we compare the physical properties (namely \mstar, SFR, \mdust; see Section \ref{sec:data} for details) of the 33 Seyfert 2 galaxies with the control sample.
In Fig. \ref{fig:comparison_cs} we show the distribution of the AGN sample and the control sample in the SFR-\mstar\ (top) and \mdust-\mstar\ (bottom) diagrams.
To highlight hidden trends that may differentiate AGN from SFGs, in the right panels of Fig. \ref{fig:comparison_cs} we present the binned version of the scaling relation presented in the left panels.
Bins were chosen to contain the same number of objects; the mean and the error bars (representing the standard deviation measured within the single bin) were obtained with bootstrap procedures using 10000 iterations.
The \mstar\ distribution for the Seyfert 2 galaxies deviates ($p=0.02$) from that of the control sample due to a gap of objects at intermediate masses ($M_{\star}\sim 10^{10-10.5}$ M$_{\odot}$), that is likely due to the lower source statistics of the AGN sample with respect to the sample of SFGs.
When matched in the low ($M_{\star}<10^{10.5}$ M$_{\odot}$) and high-stellar mass regime ($M_{\star}>10^{10.5}$ M$_{\odot}$), AGN and SFGs have almost identical distributions ($p>0.5$).
The SFR distribution in AGN deviate from that in SFGs ($p<0.001$) due to the objects with the smallest \mstar\ which show larger SFR than SFGs ($p<0.001$).
Conversely, in the highest \mstar\ regime, the two samples populate a similar region of the SFR-\mstar\ diagram ($p=0.10$ when testing the SFR in the subsamples with $M_{\star}>10^{10.5}$ M$_{\odot}$), as can be clearly seen in the top left corner of Fig. \ref{fig:comparison_cs}.
Looking at the bottom row, the Seyfert 2 galaxies show a similar distribution of \mdust\ similar to SFGs over two orders of magnitude in \mstar\ ($p\sim0.5$). 
To summarize, the AGN sample and the control sample appear to be drawn from the same parent population, since they show similar distribution in terms of \mstar\ and \mdust, with small differences in the \mstar\ distribution that are likely arising from the different object statistics of the AGN sample.
Regarding the SF activity, the AGN are likely hosted in galaxies with similar SFR to the control sample in the more massive  regime ($M_{\star}> 10^{10.5}$ M$_{\odot}$).
Less massive objects ($M_{\star}< 10^{10.5}$ M$_{\odot}$) hosting an AGN show relatively larger SFR than SFGs.
This deviation is further discussed in Section \ref{sec:scal_rel}.

%
\subsection{\lir-\lco\ relation}
\label{sec:lir_lco}
Normal SFGs, for which a relation has been found between stellar mass and star formation (MS; e.g. \citealt{Speagle14}), are thought to follow a unique \lir-\lco\ relation at all redshifts (e.g., \citealt{Daddi10b}; \citealt{Genzel10}; \citealt{Sargent14}), suggesting a ubiquitous relation between SF activity and molecular gas reservoir in normal SFGs.
Conversely, objects with higher SF efficiency as local ULIRGs and high-redshift starburst sources ($z>1$ galaxies with SFR of many hundreds of M$_{\odot}$ yr$^{-1}$; e.g., \citealt{Puglisi17}) have higher \lir/\lco\ ratios, implying a possible bimodal SF scenario.
Here, we investigate how the sample of local AGN populates the SF-molecular gas parameter space.
At first, to avoid the systemic effects introduced by the assumption of both the CO-to-H$_2$ and the SFR-to-IR conversion factors, we investigate the aperture-corrected CO(1-0) luminosity (\lco) from the single-dish observations as a function of the IR luminosity (L$_{IR}$; Fig. \ref{fig:Lco-Lir}) derived from SED fitting.
The 33 Seyfert 2 galaxies show a \lco-\lir\ ratio similar to that of the control sample of SFGs (upper panel of Fig. \ref{fig:Lco-Lir}) over two orders of magnitudes in \lir.\\
We then fit a line to the logarithms of \lco\ and \lir\ of the Seyfert 2 galaxies of the form $\log(L^{\prime}_{\rm CO(1-0)})=\alpha \log(L_{IR})+\beta$, using the ``emcee" package, a pure-Python implementation of Goodman \& Weare's Affine Invariant Markov Chain Monte Carlo (MCMC) Ensemble sampler \citep{emcee}.
The best-fit parameters of the \lir-\lco\ relation for the sample of Seyfert 2 galaxies (shown as black solid line in the bottom panel of Fig. \ref{fig:Lco-Lir}) are consistent with the trends observed in the literature for local and intermediate-redshift SFGs (e.g., \citealt{Daddi10b}; \citealt{Genzel10}; \citealt{Sargent14}).
By comparing the AGN and the control sample, they populate common regions of the \lir-\lco\ diagram, and the \lco\ of the brightest AGN in the IR in the AGN sample in particular (L$^{SF}_{IR}>10^{10.5}$ L$_{\odot}$) are statistically indistinguishable ($p>0.5$) from the SFGs.
This is in agreement with what is observed in the upper row of Fig. \ref{fig:comparison_cs}, where the AGN populate a common SFR-$M_{\star}$ plane, at least for the $M_{\star}>10^{10.5}$ M$_{\odot}$ ($p=0.1$).
%
%
\begin{figure*}[pt]
	\includegraphics[width = \textwidth,keepaspectratio=True]{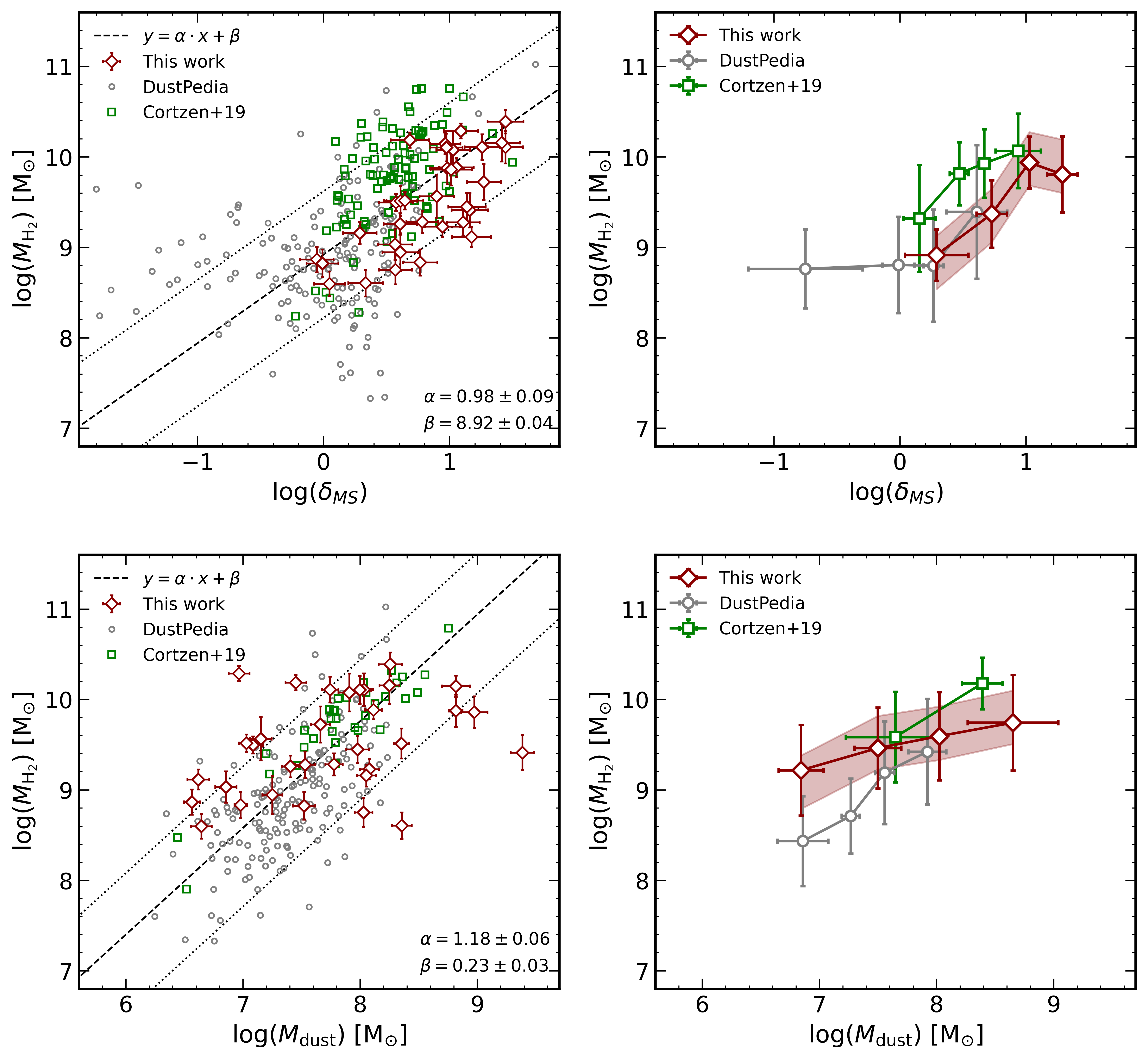}\\
	\caption{Left panels: molecular gas mass ($M_{H_2}$) versus the offset from the MS (derived assuming the relation by \citealt{Sargent14}; top panel), and host-galaxy dust mass (\mdust; bottom panel). 
	Seyfert 2 galaxies are shown with red diamonds, while the control sample are represented by gray circles (DustPedia) and green squares (Co19).
	In the bottom panel, the $M_{H_2}$-$M_{dust}$ diagram, where the $M_{H_2}$ of the 33 5MUSES SFGs were derived from the CO luminosity (Co19), assuming a \aco=4.3 \uaco.
	We do not represent the entire 5MUSES SFGs since their \mh\ are derived from the \mdust\, assuming a dust-to-gas ratio.
	For pure illustrative purposes, the best-fit trend (black-dashed line) for the control sample of inactive galaxies is shown, while the two black-dotted lines are the same trend shifted by a factor of 0.5~dex in either direction.
	The best-fit slope ($\alpha$) and normalization ($\beta$) are reported in each panel.
	Right column: same scaling relations presented on the left, but with binned values (using the same method as in Fig. \ref{fig:scal_rel}).
	The red-shaded regions are as described in Fig. \ref{fig:scal_rel}.}
	\label{fig:mdust}
\end{figure*}

\subsection{Molecular gas mass scaling relations}
\label{sec:scal_rel}
The molecular gas masses for the 33 Seyfert galaxies are presented in Table \ref{table:CO}, following the recipe described in Section \ref{sec:Mgas}.
We remind that the best estimate of \mh\ are obtained with the prescription by \cite{Narayanan12} for the \aco\ conversion factor; we also graphically present the range of \mh\ (red-shaded area in the right panels of Figs. \ref{fig:scal_rel} and \ref{fig:mdust}.) that can be obtained by assuming different CO-to-H2 conversion factors in the range between the Milky-way like value (\aco=4.3\uaco) and \aco=1.1\uaco, typically used for the central region of local AGN (e.g., \citealt{Pozzi17}; \citealt{Rosario18}).\\
To assess if, and to what extent, AGN can affect the host-galaxy molecular gas content and SF activity, we compare the molecular gas masses (\mh) as a function of different host-galaxy properties -- namely the stellar mass ($M_{\star}$) and the SFR, dust mass (\mdust) and offset from the MS ($\delta_{MS}$) -- in the AGN sample, and the control sample (Figs. \ref{fig:scal_rel} and \ref{fig:mdust}, respectively).
Looking at the top left panel of Fig. \ref{fig:scal_rel}, it is clear that the \mh\ distribution of the AGN sample deviate significantly ($p\sim0.01$) from that of normal SFGs if we compare the two samples over the two orders of magnitude in \mstar.
The low p-value observed over the entire \mstar\ range is likely due to the low \mstar\ regime ($M_{\star}<10^{10.5}$ M$_{\odot}$) where AGN show larger \mh\ with respect to SFGs.
Conversely, for more massive objects ($M_{\star}>10^{10.5}$ M$_{\odot}$), we cannot reject the hypothesis that the two distributions are similar ($p\sim0.6$).
This trend is also present when we compare the molecular gas fraction ($f_{mol}=M_{H_2}/M_{\star}$) as a function of \mstar\ with the less massive Seyfert 2 galaxies showing relatively larger $f_{mol}$ than the SFGs ($p<0.02$).\\
This discrepancy between the low and high \mstar\ ends can be justified by considering again the SFR-\mstar\ diagram in the top row of Fig. \ref{fig:comparison_cs}.
Since the SF activity is tightly related to the amount of molecular gas available to form new stars, in the low-\mstar\ regime we expect to observe larger SFR in the same AGN that showed larger \mh\ than SFGs.\\
To further test the correlation between SF and molecular gas mass, in the bottom row of Fig. \ref{fig:scal_rel}, the molecular gas masses are presented as a function of SFR.
The two quantities show a tighter correlation with respect to the \mh-\mstar\ distribution, and in agreement with the \lir-\lco\ relation discussed in Sec. \ref{sec:lir_lco}.
In the sample of 33 Seyfert 2 galaxies we do not have objects with SFR<10 M$_{\odot}$ yr$^{-1}$ - that is likely due to the 12MGS being an IR selected sample, but they match the distribution of \mh\ of normal galaxies with SFR>10 M$_{\odot}$ yr$^{-1}$ ($p\sim0.25$), as clearly visible in the lower right corner of Fig. \ref{fig:scal_rel}, where the binned values of \mh\ and SFR are shown.
By assuming a lower \aco\ value (e.g., \aco=1.1\uaco), the KS test produced a low probability ($p<0.001$), suggesting a statistical difference between the distribution of AGN and SFGs with SFR>10 M$_{\odot}$ yr$^{-1}$.\\
%
In the upper left corner of Fig. \ref{fig:mdust}, the molecular gas content as a function of the offset from the MS ($\delta_{MS}$) is shown.
The excess (or deficiency) of specific SFR (sSFR=SFR/\mstar) with respect to that expected for MS galaxies can be expressed as $\delta_{MS}=sSFR/sSFR_{MS}$.
The normalization of the MS was derived from the relation by \cite{Sargent14}, which provides the sSFR for MS galaxies, assuming the stellar mass (\mstar) and redshift of the sample of Seyfert 2 galaxies.
On average, AGN show larger offsets from the MS with respect to the normal galaxies in the control sample, as also visible in the top right panel of Fig. \ref{fig:mdust}.
This is consistent with what observed in high-redshift ($z\sim$1-3) AGN and obscured quasars (e.g., \citealt{Kakkad17}; \citealt{Brusa18}), which share similar offset from the MS to local Seyfert 2 galaxies.
The distribution of \mh over the entire range of $\delta_{MS}$ covered by the AGN sample is statistically different ($p<0.01$) from the one of the control sample. 
Since $\delta_{MS}$, as defined above, depends on the sSFR, it is clear that the part of Seyfert 2 galaxies with larger SFR than SFGs in the common low-\mstar regime (see top panels of Fig. \ref{fig:comparison_cs}) are more likely to populate the higher end of the $\delta_{MS}$ distribution with respect to the corresponding SFGs with similar \mstar.\\
The \mh-\mdust\ diagram is presented in the bottom panels of Fig. \ref{fig:mdust}.
In this case, we do not include the entire sample by Co19, since in that work the authors derive the molecular gas mass from \mdust\ by assuming a GDR, therefore making the two quantities proportional by definition.
Nevertheless, Co19 provided the \lco\ for 41 galaxies (out of the 95 we collected in Section \ref{sec:control_sample}), which we used to derive the corresponding \mh\ by assuming as CO-to-H$_{2}$ conversion factor \aco=4.3 \uaco.
Therefore, when discussing the \mh-\mdust\ distribution, we considered a control sample limited to 210 objects. 
Looking at the bottom panels of Fig. \ref{fig:mdust}, AGN seem to host larger molecular gas reservoir when compare to the SFGs over a wide range of dust masses, something that results in a low p-value ($p<0.001$) of the KS test.
Furthermore, the distribution of the \mh in the most massive AGN (i.e., with $M_{dust}>10^{7.5}$ M$_{\odot}$) are more statistically similar to the corresponding distribution of SFGs ($p>0.5$) with respect to the lower massive regime.
Interestingly, as the shaded region in the bottom right panel would suggest, by repeating the KS test assuming a lower \aco\ value for the AGN results in an increased statistical significance for the null hypothesis ($p>0.1$).\\
To conclude, we observe that AGN with larger \mstar\ are likely to host similar molecular gas content to SFGs.
This is in contrast with what was observed by \cite{Koss20b}, who found that a large sample of local AGN, selected in the hard X-ray band, show larger \mh\ with respect to their control sample of SFGs for $M_{\star}>10^{10.8}$ M$_{\odot}$.
In the work by \cite{Koss20b}, the \mh\ for both AGN and normal galaxies were derived with a single \aco=4.3\uaco, while here we adopted the prescription by \cite{Narayanan12} for AGN, that provides lower \aco\ factors (mean value \aco$\sim$ 3 \uaco, and $\sim$ 1 \uaco\ as standard deviation).
However, even if we assume \aco=4.3 \uaco\ for the Seyfert 2 galaxies, AGN and SFGs show similar \mh\ content only in the more massive regime in terms of \mstar\ ($p\sim0.2$).
Therefore, the different selection criterion adopted to build the samples \citep[hard X-ray selection in ][while in this work we consider IR selected objects]{Koss20b} is likely the reason driving this discrepancy.\\
The \mh\ distribution of the sample of Seyfert 2 galaxies does not differentiate from SFGs ($p=0.25$), when matched in the high-SFR regime (SFR$>$10 M$_{\odot}$ yr$^{-1}$).
These higher SFR and molecular gas content in local AGN could be linked to the nuclear activity. 
AGN may have had higher H$_2$ content than SFGs, but this molecular gas may have been used both as main fuel for SF (resulting in higher SFR) and for accretion onto the central SMBH (losing H$_2$). 
The deviation of the low-dust mass ($M_{dust}<10^{7.5}$ M$_{\odot}$) AGN from the \mh-\mdust\ relation is likely due to the large scatter that affects such relation, as observed in several studies on nearby galaxies (e.g., \citealt{Orellana17}; Ca20).
Owing to the higher SFRs, Seyfert 2 galaxies show larger offset from the MS, since $\delta_{MS}$ depends on SFR by definition.

%
\subsection{Molecular gas depletion times}
\label{sec:tdepl}
\begin{figure*}[ht]
	\includegraphics[width = \textwidth, keepaspectratio=True]{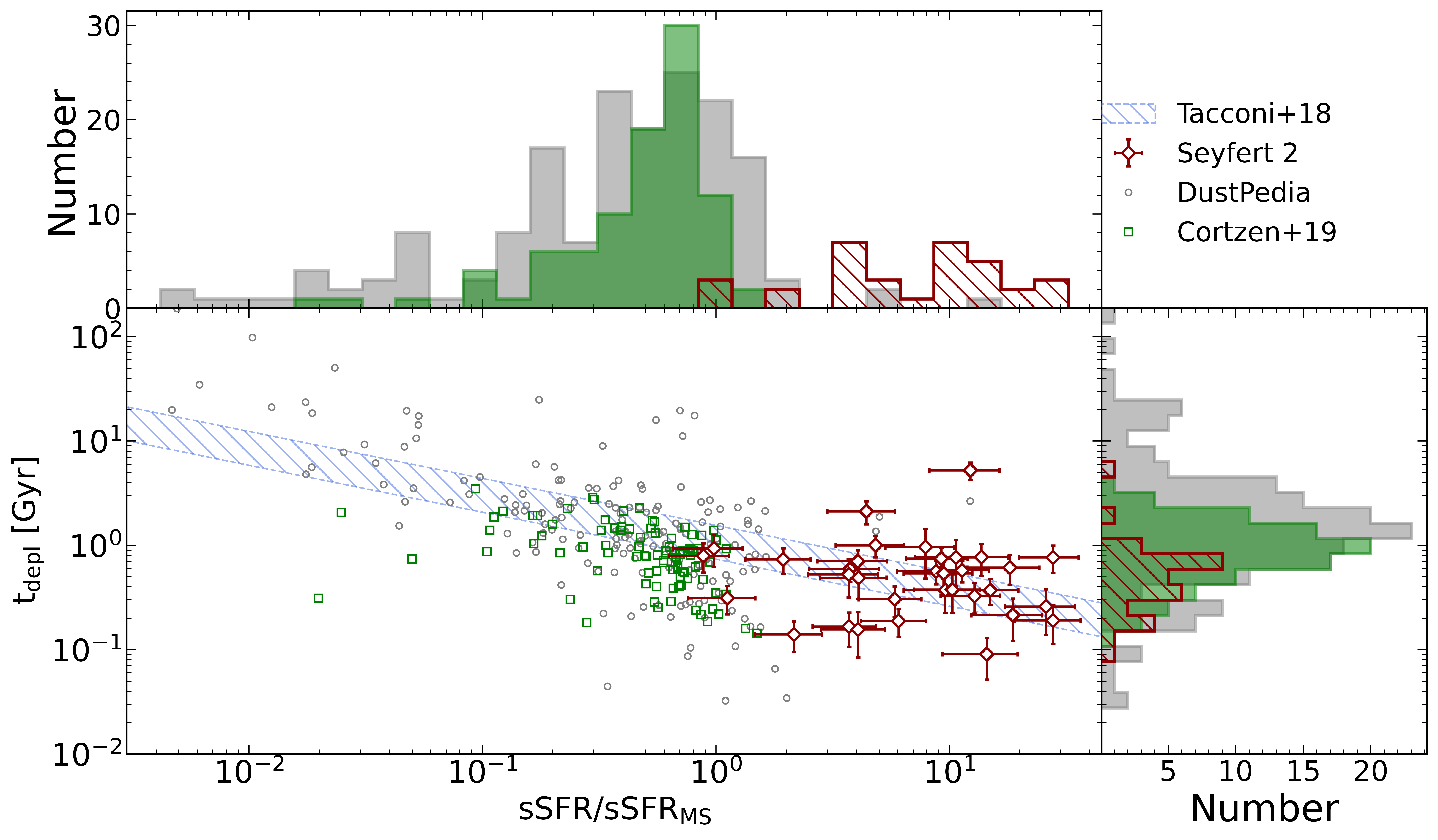}\\
	\caption{Main panel: Depletion time (in units of Gyr) vs, the distance from the MS ($\delta_{MS}$) in terms of the sSFR. 
	The Seyfert 2 galaxies are the red diamonds, while the control sample are the gray circles (DustPedia) and green squares (Co19).
	The dashed region represents the $t_{depl}$-$\delta_{MS}$ relation by \cite{Tacconi18}, in the interval of stellar masses and redshift of the Seyfert 2 galaxies.
	Upper and right panels: projected histograms of the distance from the MS and the depletion times, respectively, for the AGN sample and the control sample, following the same color coding.}
	\label{fig:tdep}
\end{figure*}

The depletion time, i.e. the ratio between the molecular gas mass and the SFR, is a widely used indicator (e.g., \citealt{Daddi10b}; \citealt{Genzel10}; \citealt{Brusa18}; \citealt{Koss20b}) of the timescale necessary for the galaxy to convert the available \mh\ into new stars at the rate of the currently ongoing SF activity.
If the AGN is able to remove part of the molecular gas, this would result in shorter \tdep\ with respect to SFGs with similar SFR.
By considering the gas mass computed in Section \ref{sec:Mgas} and the SFR provided by G16, we computed the depletion times for the sample of Seyfert 2 galaxies.
We found \tdep\ in the range $0.1<t_{depl}<7$ Gyr (with median value $\sim1$ Gyr; see the right panel of Fig. \ref{fig:tdep}), consistent with what has been reported in previous literature works in local AGN ($L_{bol}\sim10^{43-46}$ erg s$^{-1}$) with similar host-galaxy properties in terms of $M_{\star}$ and SFR (0.1$<$\tdep $<$ few Gyr; e.g., \citealt{Casasola15}; \citealt{Rosario18}; \citealt{Koss20b}).
Conversely, shorter time scales (0.01$<t_{depl}<$0.1 Gyrs) for the gas consumption have been observed in the case of high-redshift AGN and quasars (z$\sim$1-3; e.g., \citealt{Brusa18}; \citealt{Kakkad17}; \citealt{Talia18}), likely due to the combined enhancement of both SF and AGN activity at the cosmic noon \citep{MadauDickinson14}.\\
In the central panel of Fig. \ref{fig:tdep}, we plot the depletion times as a function of the offset from the MS.
In the top panel of Fig. \ref{fig:tdep}, the histogram of $\delta_{MS}$ is shown: Seyfert 2 galaxies show systematically larger distances from the MS with respect to the distribution of SFGs.
Looking at the depletion time distribution for the Seyfert 2 galaxies and the control sample (right panel of Fig. \ref{fig:tdep}), both samples peak in a similar regime (mean $t_{depl}\sim0.6$ and $1$ Gyr for AGN and control sample, respectively), but the KS test excludes that the two samples are drawn from a common distribution ($p\sim0.01$).
The low p-value is likely driven by the large tail of DustPedia galaxies with relatively long \tdep.
We also test how the distribution of \tdep\ for the AGN changes by assuming a single \aco=4.3 \uaco\ for both AGN and SFGs.
In this case, we find that the Seyfert 2 galaxies and the control sample are statistically indistinguishable ($p=0.8$), with the resulting \tdep\ distribution for AGN peaking at $t_{depl}\sim0.9$ Gyr.
Conversely, by shifting the \tdep\ for the AGN along the vertical axis, which is equivalent to assuming progressively smaller \aco\ values down to \aco=1.1\uaco, the \tdep\ distribution would consequently be shifted towards lower values, making it deviate significantly from that of SFGs ($p<0.001$), bringing them, at the same time, closer to the MS.
At the same time, the large $\delta_{MS}$ of AGN cause them to deviate from the trend observed in MS galaxies, represented by the relation by \cite{Tacconi18}, which describes the expected \tdep\ for MS galaxies, at a given \mstar\ and redshift (shown in Fig. \ref{fig:tdep} at the representative median redshift and $M_{\star}$ of the Seyfert 2 galaxies).
We conclude that the sample of local Seyfert 2 galaxies covers the shorter-\tdep\ regime of local galaxies, with $\sim$70\% of the AGN sample having $t_{depl}\sim0.3-1.0$ Gyr, while the corresponding fraction of SFGs show $t_{depl}\sim0.3-3.0$ Gyr), which makes the distributions statistically different.

%
%
\subsection{PAH as tracer of the molecular gas}
\begin{figure*}[pt]
	\includegraphics[width = \textwidth, keepaspectratio=True]{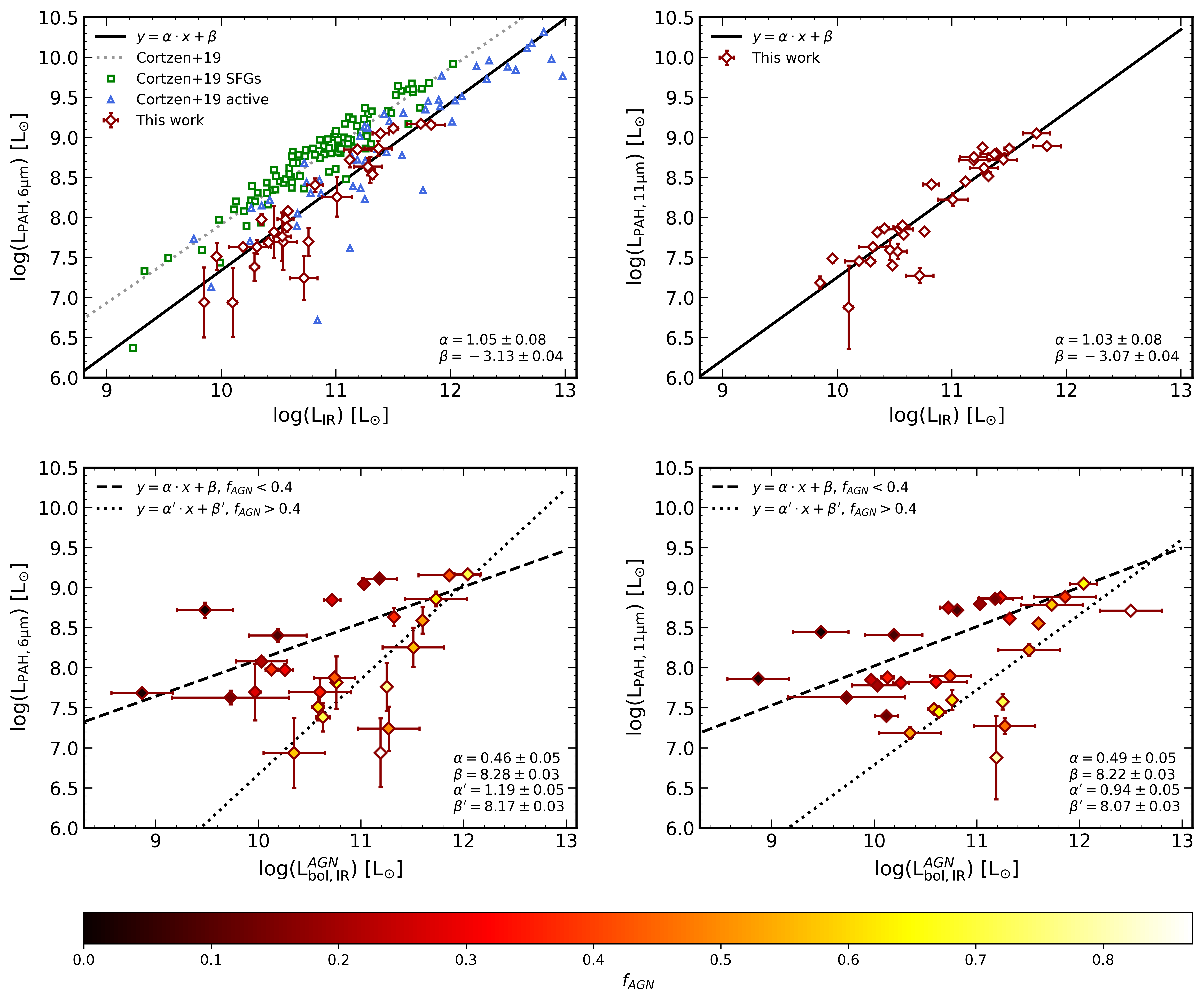}
	\caption{Top left panel: 6.2 $\mu$m PAH feature luminosity versus IR luminosity.
	Seyfert 2 galaxies are shown as red diamonds, while green squares and blue triangles indicate the inactive and active galaxies in the sample by Co19, classified based on the EW of the PAH features at 6.2 and 11.3 $\mu$m.
	The best fit relation (best fit parameters are reported in the lower-right corner) is the black solid line, while the gray-dotted line is the result of the best fit by the Co19.
	Top right panel: 11.3 $\mu$m PAH feature luminosity versus IR luminosity for the Seyfert 2 galaxies.
	Points and lines are coded as in the left panel.
	Bottom panels, from left to right: 6.2 $\mu$m and 11.3 $\mu$m PAH luminosity, respectively, versus the bolometric luminosity of the AGN, derived from the broad band SED decomposition performed by G16. 
	Data are color-coded as a function of the fraction of the AGN emission with respect to the galaxy global outcome in the 5-40 $\mu$m band.
	Dashed (dotted) lines are the best fit relation for the subsample of AGN with $f_{AGN}<0.4$ ($>0.4$).
	}
	\label{fig:IR-PAH}
\end{figure*}

The \lir-$L_{PAH}$ relation has been widely used as a diagnostic to distinguish between different galaxy populations (e.g., MS, starburst, AGN; e.g., Co19; \citealt{Minsley20a}), as they tend to occupy different parts of the diagram.  
When comparing $L_{PAH}$ at a given \lir\ between AGN and SFGs, AGN usually appear as outliers of the relation, suggesting that $L_{PAH}$ can be used to assess the impact of the AGN emission in the central region of the galaxy with respect to other excitation mechanisms.
Since only the sample by Co19 has available PAH feature measurements, in this section we only compare the AGN sample to the SFGs by Co19, as shown in Fig. \ref{fig:IR-PAH}.\\
The luminosity of the MIR features provided by HC11 (i.e., the PAH at 6.2 and 11.3 $\mu$m) scale with the IR luminosity, as observed in many local objects (e.g., \citealt{Alonso-Herrero16}; \citealt{Jensen17}; \citealt{Kirkpatrick17}), even in the presence of nuclear activity.
We performed a linear fit of the \lir-\lpah\ relation, $\log(L_{PAH})=\alpha \log(L_{IR})+\beta$, for both the PAH features at 6.2 and 11.3 $\mu$m.
The results of this fit are presented in Fig. \ref{fig:IR-PAH}.
We found a slope ($\alpha=1.05\pm0.02$) slightly steeper than that of Co19, and a significantly lower normalization ($\beta^{\rm Cortzen+19}-\beta^{\rm this\ work}\sim0.6$~dex).
The sample of active galaxies shows lower 6.2 $\mu$m PAH luminosity for a given \lir, suggesting a potential effect of the nuclear activity on the emission of this MIR feature.
Furthermore, to test the consistency of the $L_{PAH, 6.2\mu m}$ deficiency in galaxies with an ongoing nuclear activity, in the top left panel of Fig. \ref{fig:IR-PAH} we included the active objects from  Co19 (i.e., 61 AGN, or composite, with EW$_{6.2\mu m}<0.4 \mu$m; see also \citealt{Sargsyan12}; \citealt{Stierwalt14}), which were previously excluded on the basis of the EW selection criteria.
Indeed, lower $L_{PAH}$-\lir\ ratios have been reported in active galaxies compared to SFGs (e.g., \citealt{Armus07}; \citealt{Valiante07}; \citealt{Sajina08}; \citealt{Diamond-Stanic10}), suggesting that the strong radiation field produced by the AGN can, at least in part, destroy part of PAH molecules rather than exciting them.
\\
We then found similar results for the fit of $L_{PAH, 11.3\mu m}$-\lir\ ($\alpha=1.03\pm0.03$; see the right panel of Fig. \ref{fig:IR-PAH}).
This is in agreement with the literature, since 6 $\mu$m feature is likely excited by SF-related emission, while the feature at longer wavelength could be more affected by the presence of AGN, whose dust reprocessed emission peaks in the 10-30 $\mu$m regime (e.g., \citealt{Mullaney11}).
The depleted PAH luminosity in the presence of nuclear activity is consistent with the recent results by \cite{GarciaBernete21}, where the authors observed depleted PAH emission in the nuclear region of local AGN, while at larger distance from the nucleus AGN-dominated objects show similar PAH luminosity to SFGs.
Thus, the extrapolation of the molecular gas mass from the PAH feature luminosity (e.g., as suggested by Co19) can induce a significant underestimation of \mh\ if the presence of nuclear activity has not been properly identified, as it happens in the case of heavy extinction of weak nuclear emission, as in the case of some local extremely obscured Seyfert 2 galaxies (e.g., \citealt{Marchesi18}).\\
To further investigate how the presence of nuclear activity may affect L$_{PAH}$, in the lower panels of Fig. \ref{fig:IR-PAH} we plot the 6.2 $\mu$m and 11.3 $\mu$m PAH luminosity, respectively, versus the bolometric luminosity of the AGN ($L_{bol, IR}^{AGN}$), provided by G16.
Following the results by G16 (see Fig. 10 in G16), we divided the sample in two subsamples on the basis of the relative contribution of the AGN (f$_{AGN}$) to the global outcome of the source in the 5-40 $\mu$m band, and we fitted separately the two subsamples. 
We found that AGN with a relatively larger contribution of the AGN to the MIR outcome of the galaxy \(f_{AGN}>0.4\) show fainter PAH emission (both the 6.2 $\mu$m and 11.3 $\mu$m features) for a given AGN bolometric luminosity.
The fainter L$_{PAH}$ for larger contribution from the AGN dust reprocessed emission ($f_{AGN}$) at a given AGN bolometric luminosity supports the negative effect of nuclear activity on the emission from PAH molecules.
This trend is clearly visible for both the emission from 6.2 and 11.3 $\mu$m features, which arise primarily from ionized and neutral PAH molecules, respectively, suggesting that the AGN emission affects both molecular phases (we refer to \citealt{Li20} for more details).
\cite{GarciaBernete21} found higher 11.3-to-6.2 $\mu$m luminosity ratio in AGN with respect to SFGs, which is in contrast with our results, but it is likely due to the large presence ($\sim$56\%) of Seyfert 1 objects in their sample of AGN.
We conclude that the impact of AGN emission on the ISM is clearly visible in the case of PAH emission, when the AGN is able to reduce the emission from molecules or destroying them.
In conclusion, it is worth mentioning that \cite{Tommasin10} attributed at least part of the depleted PAH emission in active galaxies to the stronger AGN continuum which is a relevant contribution to the MIR emission in galaxies.
However, the authors in \cite{Tommasin10} suggested that this effect is more prominent in unobscured Seyfert galaxies (i.e. type 1), while stacked MIR spectra of Seyfert 2 objects resemble those of non-Seyfert's and starburst sources.   
In the future, to definitely assess the impact of the AGN emission on the ISM, as well as its ability to impact the SF activity, we need high-spatially resolved observations (e.g., using ALMA to trace the CO emission down to the giant molecular cloud scale, $\sim50-100$ pc, and future JWST observation for the MIR features) to study the properties and the physical condition of the ISM.
%
\section{Summary and conclusions}
\label{sec:conclusion}
In this work we studied the properties of a sample of 33 local Seyfert 2 galaxies with the aim of understanding the impact of the nuclear activity on the host-galaxy molecular gas content and SF activity.
By considering new and archival CO spectroscopic data, we estimated the mass of the molecular component of the ISM -- the key ingredient to form new stars.
We exploited the results of the detailed SED decomposition by G16, providing the characterization of each galaxy, in terms of \mstar, SFR, \lir, and emission from the AGN.   
By comparing the molecular gas content ($M_{H_2}$) and the relative depletion time in the sample of local Seyfert 2 galaxies matched in different host-galaxy properties (\mstar, SFR, \mdust) to a control sample of SFGs, we investigated if the nuclear activity affect the host-galaxy molecular gas reservoir. 
Furthermore, we investigated the effect of the nuclear activity onto the PAH features, which are widely used as tracers of SF activity, while the presence of the AGN can actually affect their luminosity -- AGN have been observed both suppressing and enhancing the PAH emission (e.g., \citealt{Sajina08} and \citealt{Jensen17}, respectively)\\
The main results of this work are as follows:
\begin{itemize}
    \item Aperture corrected molecular gas masses for 33 objects have been provided, derived by converting low-J CO luminosity obtained from new and archival single-dish observations.
    \item The Seyfert 2 galaxies with $M_{\star}<10^{10.5}$ M$_{\odot}$ show larger molecular gas mass and fraction compared to SFGs (with similar \mstar), while they cannot be distinguished from SFGs in terms of the molecular gas content at larger \mstar.
    This is likely due to the presence of more actively star-forming galaxies in low \mstar\ regime in the AGN sample with respect to the control sample. 
    \item AGN show similar \mh\ content to SFGs when the two samples are matched in SFR.
    The AGN which deviate most from the \mh-\mstar\ trend above discussed are likely hosted in galaxies which are forming new stars at higher rate (SFR$\sim$5 M$_{\odot}$ yr$^{-1}$) than the control-sample galaxies in the same low \mstar\ regime, due to the availability of larger molecular gas reservoir.
    The tighter correlation we observed between SFR and \mh\ reflects the physical correlation between the ongoing SF activity and the available molecular gas reservoir to form new stars.
    \item When comparing the depletion times between the Seyfert 2 galaxies and the control sample, AGN lie in the short-\tdep\ regime ($\sim$0.3-1 Gyr) with respect to those observed in SFGs.
    However, the observed range of \tdep\ in the Seyfert 2 sample is consistent with values expected for both local AGN and normal galaxies, which confirms that the local AGN do not significantly reduce the global molecular gas content of the host galaxy.
    The Seyfert 2 galaxies show significantly larger distance from the MS, as parametrized in terms of the sSFR, irrespective of \tdep.
    This effect is driven by the higher SFR observed in the low \mstar\ regime of AGN with respect to SFGs. 
     \item By studying the \lpah-\lir\ scaling relations with the PAH features at 6.2 and 11.3 $\mu$m, we found that the sample of local Seyfert 2 galaxies shows lower \lpah/\lir\ ratios with respect to SFGs, as previously observed in AGN-dominated samples.
     Furthermore, when studying the PAH emission as a function of the AGN bolometric luminosity, we found that the objects with a relatively larger contribution from the AGN in the MIR \(f_{AGN}\) show fainter PAH emission.
     This suggests that in obscured AGN, the nuclear activity is able to suppress, at least in part, the emission of PAH features, being more efficient when the AGN emission dominate the MIR regime.
\end{itemize}
Finally, we found no clear evidence that local AGN can significantly reduce the global molecular gas reservoir of their host galaxy.
This is consistent with the depleted emission from the molecular gas observed in recent high-spatially resolved studies of the nuclear regions of local Seyfert galaxies (e.g., \citealt{GarciaBurillo21}), and in AGN-dominated regions of sub-galactic scales (few kpc; e.g., \citealt{Ellison21}).\\
The future scope of this work will include multi-wavelength high-spatially resolved data to investigate the properties and physical conditions of the molecular gas component in the central region of local Seyfert galaxies.
Namely, by combining interferometric observations (such ALMA and NOEMA) to trace the molecular gas, with optical data (MUSE/VLT) to determine the ISM metallicity (e.g., see \cite{Kreckel19}), as well as the advent of JWST, providing spatially resolved observations of the MIR features, we will be finally able to assess the effect of AGN activity on the ISM at increasing distance from the nucleus. 
In this regard, the IR-selected sample of local Seyfert galaxies presented here (up to the 30\% of the sample presented by G16 have interferometric observations) represents an ideal reference sample to understand the interplay between SF and AGN activity. 

%
%
\begin{acknowledgements}
We are grateful to the anonymous referee for her/his constructive comments and suggestions, which helped improving the quality of the paper.
Grateful thanks to M. Talia and A. Giannetti for his important contribution to design the proposal for the APEX observations (project 0103.F-9311, PI: Salvestrini) which have been used in this work.
We are grateful to A. Feltre and L. Marchetti for providing us with SALT optical emission lines measurements for two galaxies in our sample prior to publication (projects: 2018-1-SCI-029 and 2020-2-MLT-006, PI: L. Marchetti; Feltre et al., in prep.).
The author is thankful to ESO for providing him valuable hospitality for a five-months visiting period, during which part of this work was pursued.
F. S., V. C. and F. P. acknowledge funding from the INAF mainstream 2018 program ''Gas-DustPedia: A definitive view of the ISM in the Local Universe``.
S.M. acknowledges funding from the the INAF ''Progetti di Ricerca di Rilevante Interesse Nazionale`` (PRIN), Bando 2019 (project: ''Piercing through the clouds: a multi-wavelength study of obscured accretion in nearby supermassive black holes``).
S.A., gratefully acknowledges support from an ERC Advanced Grant 789410.
Based on observations collected at the European Southern Observatory under ESO programme 0103.F-9311(A) with the Atacama Pathfinder EXperiment (APEX) telescope. APEX is a collaboration between the Max Planck Institute for Radio Astronomy, the European Southern Observatory, and the Onsala Space Observatory. Swedish observations on APEX are supported through Swedish Research Council grant No 2017-00648. The time granted was used to obtain data for the target of this work.
This research has made use of the NASA/IPAC Extragalactic Database (NED), which is funded by the National Aeronautics and Space Administration and operated by the California Institute of Technology.
We acknowledge the usage of the HyperLeda database (\url{http://leda.univ-lyon1.fr})
\end{acknowledgements}
%
%
\bibliographystyle{aa}
\bibliography{Mmol_salvestrini.bib}

%
%
\onecolumn
\begin{appendix}
%
%
\section{Kolmogorov-Smirnov tests}
\label{app:KS-test}
Here we show the histograms of the p-value distributions obtained by repeating the KS test on the 1000 copies of the AGN and SFGs samples for each of their properties, as described in Section \ref{sec:results}.
The galaxy property for which we tested the potential differences in the distribution between the AGN and the control sample is reported in the title of each panel.
We also tested  the distribution of a given quantity (e.g., \mh) in two different regimes for another physical quantity (e.g., \mstar) by splitting the sample on the base of a given threshold, e.g. in the central row of Fig. \ref{fig:all_histo_p2} the \mh\ distributions of AGN and SFGs having smaller (left panel) and larger (right panel) \mstar\ than $10^{10.5}$ M$_{\odot}$ is shown.\\ 
\begin{figure*}[hbp]
	\includegraphics[width = \textwidth, keepaspectratio=True]{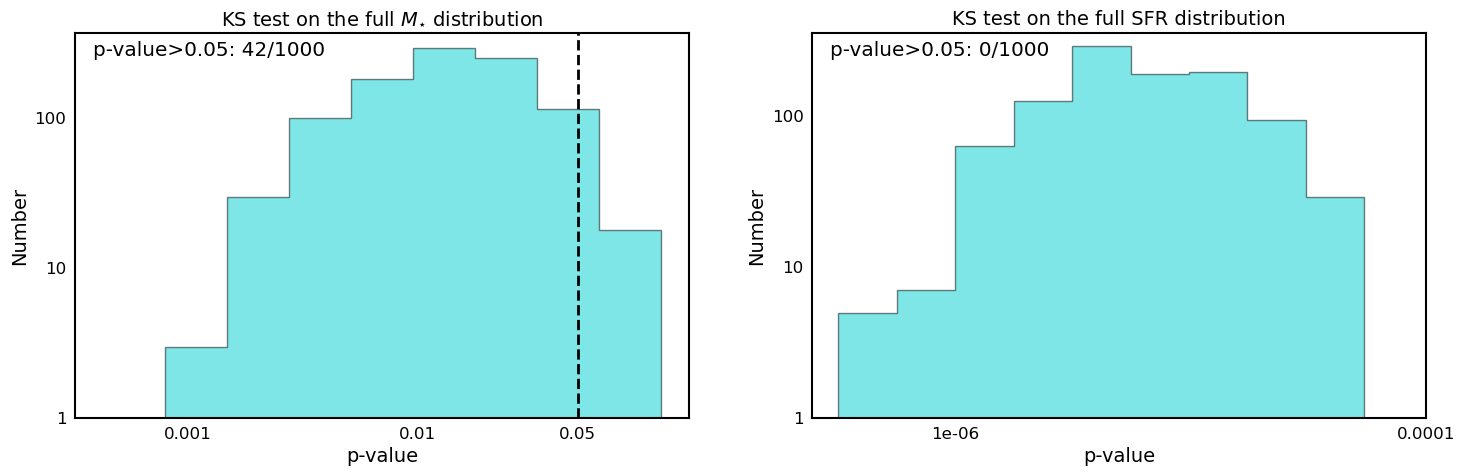}
	\includegraphics[width = \textwidth, keepaspectratio=True]{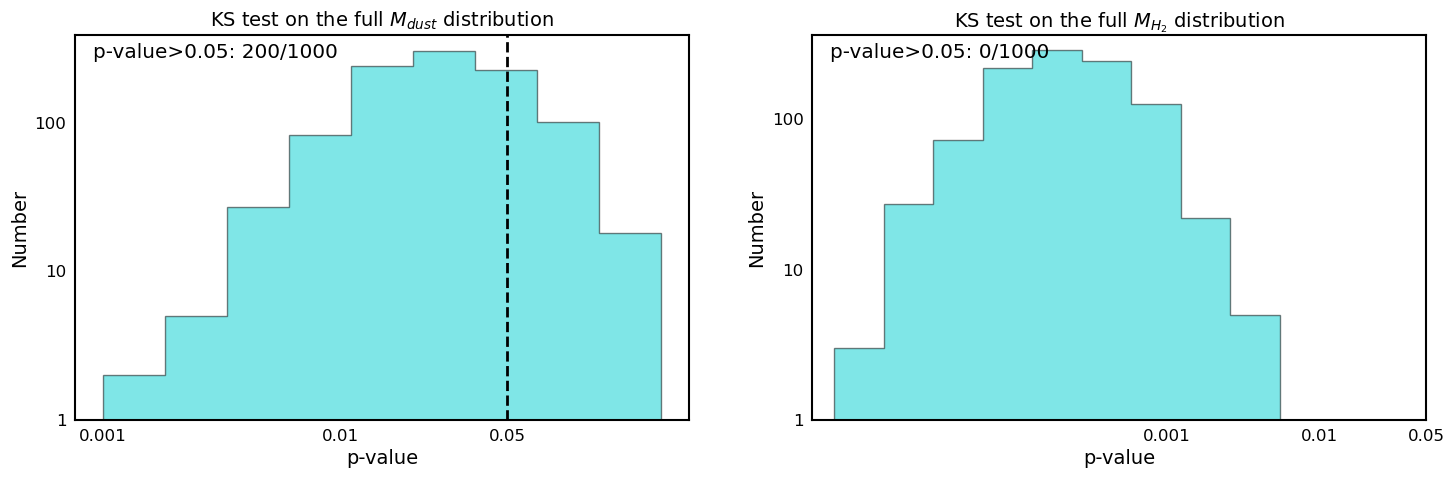}
	\includegraphics[width = \textwidth, keepaspectratio=True]{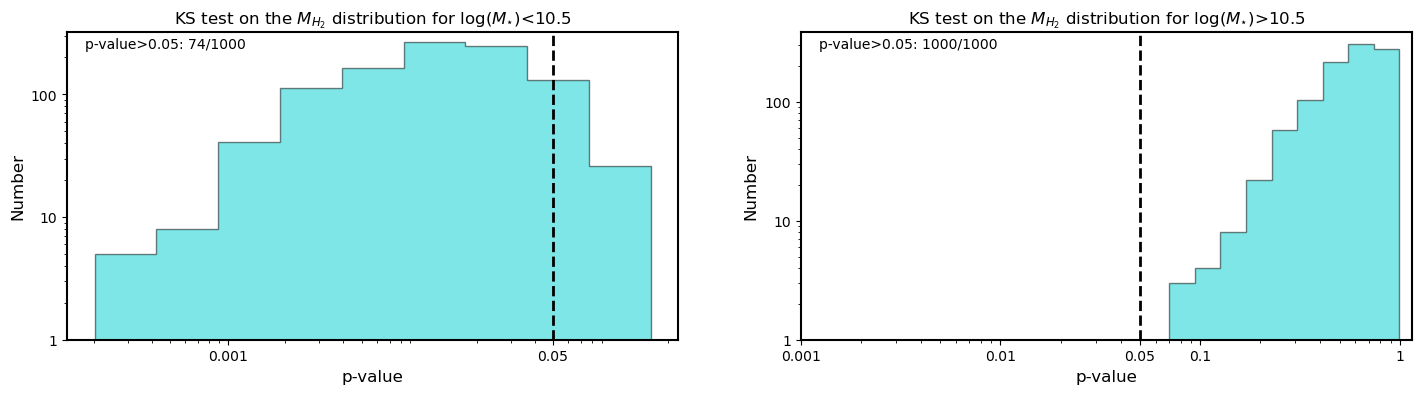}
	\caption{Upper panels: histograms of the p-values of the KS test performed on the \mstar\ (left panel) and SFR (right panel) simulated distributions of AGN and the control sample of SFGs.
	On the y-axis, the logarithm of the number of entries are divided in ten equally log-spaced bins, while the p-values resulting from the 1000 KS tests are reported on the x-axis.
    The threshold above which we cannot reject the null-hypothesis that the samples are drawn from the same distribution is represented by a vertical, black-dashed line at $p=0.05$.
	Central panels: same as the upper row with \mdust\ (left panel) and \mh\ (right panel) distributions.
	Bottom panels: results of the KS tests on the \mh\ distribution in the $log(M_{\star})<10.5$ M$_{\odot}$ (left panel) and $log(M_{\star})>10.5$ M$_{\odot}$ (right panel) regimes.
	}
	\label{fig:all_histo_p1}
\end{figure*}
\begin{figure*}[ht] 
\ContinuedFloat
	\includegraphics[width = 0.5\textwidth, keepaspectratio=True]{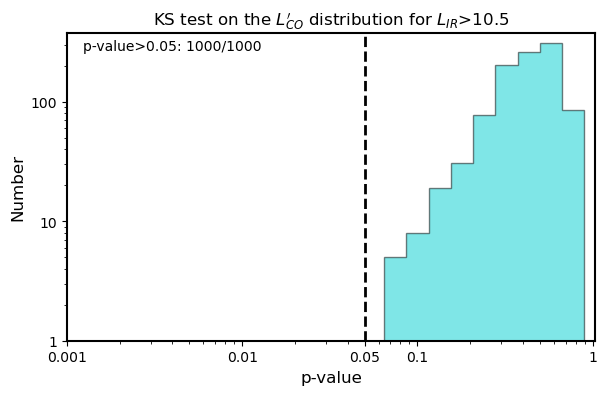}	
	\includegraphics[width = 0.5\textwidth, keepaspectratio=True]{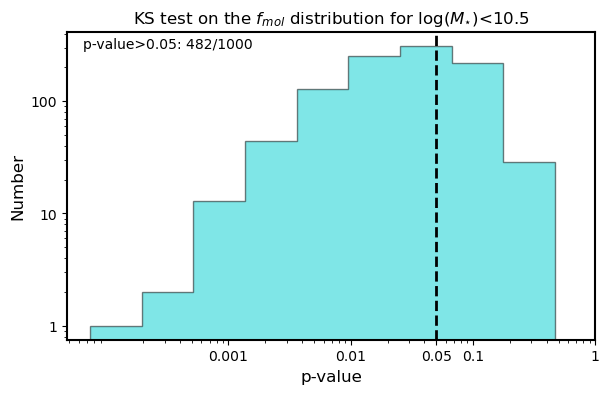}\\
	\includegraphics[width = 0.5\textwidth, keepaspectratio=True]{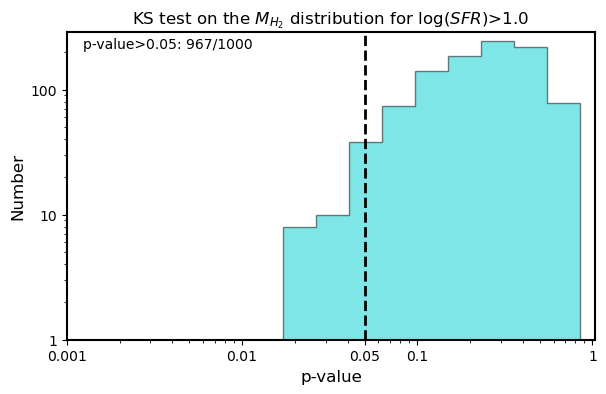}
	\includegraphics[width = 0.5\textwidth, keepaspectratio=True]{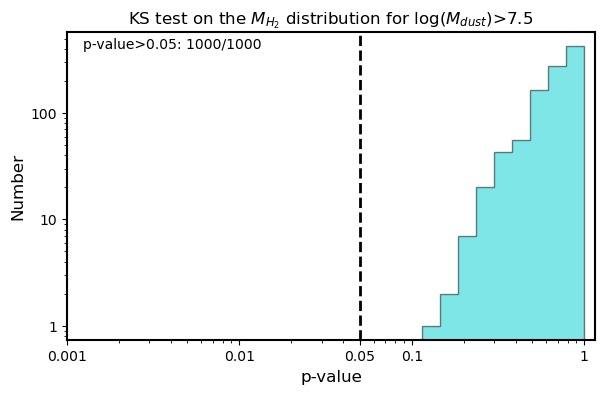}
	\caption{Continued, top panels: The distribution of p-values obtained by comparing \lco\ in the $log(L_{IR})>10.5$ L$_{\odot}$ regime (left panel), and $f_{mol}$ in the $log(M_{\star})<10.5$ M$_{\odot}$ regime (right panel).
	Bottom panels: The KS test of the \mh\ distribution for the AGN and SFGs matched in the $log(SFR)>1$ M$_{\odot}$ yr$^{-1}$ (left panel), and $log(M_{dust})>7.5$ M$_{\odot}$ (right panel) regimes, respectively.}
	\label{fig:all_histo_p2}
\end{figure*}

\end{appendix}

\end{document}